\documentclass[lettersize,journal]{IEEEtran}

\IEEEoverridecommandlockouts
\usepackage{cite}
\usepackage{amsmath,amssymb,amsfonts}
\usepackage{algorithmic}
\usepackage{graphicx}
\usepackage{textcomp}
\usepackage{xcolor}
\def\BibTeX{{\rm B\kern-.05em{\sc i\kern-.025em b}\kern-.08em
		T\kern-.1667em\lower.7ex\hbox{E}\kern-.125emX}}
	
\makeatletter
\def\bstctlcite{\@ifnextchar[{\@bstctlcite}{\@bstctlcite[@auxout]}}
\def\@bstctlcite[#1]#2{\@bsphack
	\@for\@citeb:=#2\do{%
		\edef\@citeb{\expandafter\@firstofone\@citeb}%
		\if@filesw\immediate\write\csname #1\endcsname{\string\citation{\@citeb}}\fi}%
	\@esphack}
\makeatother

\usepackage[ruled,vlined]{algorithm2e}    
\usepackage{setspace}
\usepackage{algorithmic}
\usepackage{multirow}
\usepackage{comment}
\usepackage{tabularx}
\usepackage{booktabs}
\usepackage{caption}
\usepackage{subcaption}
\usepackage[colorinlistoftodos]{todonotes}
\usepackage{stfloats}

\newcommand\norm[1]{\left\lVert#1\right\rVert}
\newcommand{\RomanNumeralCaps}[1]
{\MakeUppercase{\romannumeral #1}}

\newcolumntype{L}{>{\centering\arraybackslash}m{4.5cm}}
\newcolumntype{K}{>{\centering\arraybackslash}m{2cm}}
\newcolumntype{R}{>{\centering\arraybackslash}m{4.5cm}}

\usepackage{scalerel,stackengine}
\stackMath
\newcommand\reallywidehat[1]{%
	\savestack{\tmpbox}{\stretchto{%
			\scaleto{%
				\scalerel*[\widthof{\ensuremath{#1}}]{\kern-.6pt\bigwedge\kern-.6pt}%
				{\rule[-\textheight/2]{1ex}{\textheight}}
			}{\textheight}%
		}{0.5ex}}%
	\stackon[1pt]{#1}{\tmpbox}%
}
\usepackage{color, colortbl}
\definecolor{Gray}{gray}{0.93}
\definecolor{Gray2}{gray}{0.8}
\definecolor{LightCyan}{rgb}{0.88,1,1}

\begin{document}
	\bstctlcite{IEEEexample:BSTcontrol}
	
	\title{Direct Link Interference Suppression for Bistatic Backscatter Communication in Distributed MIMO
		
		\thanks{
			Ahmet Kaplan and Erik G. Larsson were supported by the REINDEER project of the European Union’s Horizon 2020 research and innovation program under grant agreement No.	101013425, and in part by ELLIIT and the Knut and Alice Wallenberg (KAW) Foundation. Preliminary results of this article were presented at the 2022 IEEE Globecom \cite{kaplan2022direct}.	
			
			Ahmet Kaplan and Erik G. Larsson are with the Department of
			Electrical Engineering (ISY), Linköping University, 58183 Linköping, Sweden
			(e-mail: ahmet.kaplan@liu.se; erik.g.larsson@liu.se).
			
			Joao Vieira is with Ericsson Research, 22362 Lund, Sweden (e-mail: joao.vieira@ericsson.com).

			©2023 IEEE. Personal use of this material is permitted. Permission from IEEE must be obtained for all other uses, in any current or future media, including reprinting/republishing this material for advertising or promotional	purposes, creating new collective works, for resale or redistribution to servers
			or lists, or reuse of any copyrighted component of this work in other works.
			
			This paper will appear in the IEEE Transactions on Wireless Communications, 2023.
			
			Digital Object Identifier (DOI): 10.1109/TWC.2023.3285250
			
		}
	}
	
	\author{Ahmet Kaplan,~\IEEEmembership{Graduate Student Member,~IEEE}, Joao Vieira, Erik G. Larsson,~\IEEEmembership{Fellow,~IEEE}}

\maketitle
	
	\begin{abstract}
		Backscatter communication (BC) is a promising technique for future Internet-of-Things (IoT) owing to its low complexity, low cost, and potential for energy-efficient operation in sensor networks. 
		There are several network infrastructure setups that can be used for BC with IoT nodes. 
		One of them is the bistatic  setup where typically there is a need for high dynamic range and high-resolution analog-to-digital converters at the reader. 
		In this paper, we investigate a bistatic BC setup with multiple antennas. 
		We propose a novel transmission scheme, which includes a protocol for channel estimation at the carrier emitter (CE) as well as a transmit beamformer construction that suppresses the direct link interference between the two ends of a bistatic link (namely CE and reader), and increases the detection performance of the backscatter device (BD) symbol. 
		Further, we derive a generalized log-likelihood ratio test (GLRT) to detect the symbol/presence of the BD.
		We also provide an iterative algorithm to estimate the unknown parameters in the GLRT. 
		Finally, simulation results show that the required dynamic range of the system is significantly decreased, and the detection performance of the BD symbol is increased, by the proposed algorithm compared to a system not using beamforming at the CE.
	\end{abstract}
	
	
	\section{Introduction}
	There were almost 15 billion Internet-of-Things (IoT) connections in 2021, and this number   is expected  to reach 30 billion by 2027 \cite{cerwall2021ericsson}. Backscatter communication (BC) is a promising technique for massive connectivity in 6G \cite{chowdhury20206g} owing to its low cost, low complexity, and potential to enable energy-efficient solutions for battery-less IoT devices and sensors. 
	
	In a BC setup, we have the following types of equipment: a carrier emitter (CE), a reader, and a backscatter device (BD). \footnote{A variety of CE and reader devices have been considered in the literature, including but not limited to base stations, Wi-Fi access points, distributed antenna panels, user devices such as smartphones and smartwatches, and software-defined radios \cite{kaplan2022direct, mishra2019optimal, iyer2016inter,varshney2017lorea,zhang2016enabling}.}
	The main BC configurations are
	\begin{itemize}
		\item \textbf{Monostatic:} In a monostatic BC (MoBC) setup, the CE and reader are co-located and share parts of the same infrastructure. 
		For example, the CE and reader may share the same antenna elements. 
		The CE sends a radio frequency (RF) signal to the BD, and the BD modulates the incoming RF signal and backscatters it to the reader \cite{basharat2021reconfigurable}. 
		A monostatic system suffers from round-trip path loss, and 
		requires full-duplex technology if the same antennas are simultaneously used for transmission and reception.
		
		\item \textbf{Bistatic:} In a bistatic BC (BiBC) setup, the CE and reader are spatially separated from each other, and therefore do not share RF circuitry, which is beneficial for many reasons \cite{kimionis2014increased}. For example, the CE and reader can be located to decrease the round-trip path loss. 
		
		\item \textbf{Ambient:} In an ambient BC (AmBC) system, the CE and reader are separated in a similar way as in a bistatic system. 
		However, the ambient system does not have a dedicated CE but instead relies on ambient RF sources such as Bluetooth, Wi-Fi, or TV signals \cite{basharat2021reconfigurable}.
	\end{itemize}
	
	\subsection{Motivation}	
	The focus of this paper is a BiBC setup with a multiple-antenna CE and multiple-antenna reader corresponding to a distributed MIMO setup \cite{van2019radioweaves} with one antenna panel functioning as CE and one as reader.
	In BiBC, due to the double path-loss effect on the two-way backscatter link, the received backscattered signal is typically weak compared to the direct link interference (DLI) from a CE.
	This requires a high dynamic range of the circuitry in the reader: this dynamic range must be  proportional to the signal strength ratio between the weak backscattered signal and the received signal from the direct link \cite{biswas2021direct}.
	As a result, a high-resolution analog-to-digital converter (ADC) is required to detect the weak backscattered signal under heavy DLI; this is an important consideration as with multiple-antenna technology,  ADCs are  major power consumers \cite{mollen2016uplink}. 
	Moreover, the backscattered signal is pushed to the last bits of ADC due to the DLI which causes a low signal-to-interference-plus-noise ratio (SINR) \cite{biswas2021direct}. 
	In this paper, we propose a new transmission scheme that reduces the DLI, along with a detection algorithm for use at the reader; see Section~\ref{sec:contributions} for specifics.
	
	\subsection{Review of Previous and Related Work}
		
	In this subsection, we  review the literature on BC with multiple-antenna technology and the literature on interference suppression for BC.  Table \ref{tab:Literature}   provides an overview of this literature, and places our contribution in context.
	
	\begin{table*}[tbp]
		\caption{Overview of Related Literature on Backscatter Communication}
		\centering
		\label{tab:Literature}
		\resizebox{1\textwidth}{!}{\begin{tabular}{|c|c|c|c|c|c|c|c|c|c|c|} 		
				\hline \multirow{2}{*}{ Ref } &  \multicolumn{3}{c|}{ Setup } &  \multicolumn{2}{c|}{ CE Antenna } & \multicolumn{2}{c|}{ Reader Antenna } & \multicolumn{2}{c|}{ BD Antenna } & \multirow{2}{*}{ Interference Supression } \\ 	\cline{2-10}  
				& MoBC & BiBC & AmBC & Multi & Single & Multi & Single & Multi & Single & \\ 			
				\hline 
				\cite{stockman1948communication, zawawi2018multiuser}& $\checkmark$ & &  & & $\checkmark$ & & $\checkmark$ & & $\checkmark$ & \\
				\hline 
				\cite{he2011closed} & $\checkmark$  &  & & &  $\checkmark$ &  & $\checkmark$ & $\checkmark$ & & \\
				\hline
				\cite{mishra2019optimal,kashyap2016feasibility, liu2014multi} & $\checkmark$ & & & $\checkmark$ & & $\checkmark$ & & & $\checkmark$ & \\ 		
				\hline 
				\cite{griffin2008gains, boyer2013backscatter} & $\checkmark$ & & & $\checkmark$ & & $\checkmark$ & & $\checkmark$ & & \\ 			
				\hline 
				\cite{villame2010carrier, brauner2009novel} &$\checkmark$ &  & &  &$\checkmark$  &   &$\checkmark$ &\multicolumn{2}{c|}{N/A} & $\checkmark$ \\
				\hline 
				\cite{bharadia2015backfi} &$\checkmark$ &  & & &$\checkmark$  & & $\checkmark$ & & $\checkmark$ & $\checkmark$\\
				\hline 
				\cite{kimionis2014increased, hua2020bistatic,  basharat2021reconfigurable} & & $\checkmark$ & & & $\checkmark$  &  & $\checkmark$ & & $\checkmark$ & \\
				\hline 
				\cite{qu2022channel} & & $\checkmark$ & & $\checkmark$ &  & $\checkmark$ & & & $\checkmark$ & \\
				\hline 
				\cite{li2019capacity, tao2021novel} & & $\checkmark$ & & &  $\checkmark$ & & $\checkmark$ & & $\checkmark$ & $\checkmark$ \\
				\hline   					
				\cite{varshney2017lorea} & & $\checkmark$ & &\multicolumn{2}{c|}{Unspecified}   & & $\checkmark$&  & $\checkmark$ & $\checkmark$ \\ 		
				\hline  		
				\cite{duan2017achievable} & & $\checkmark$ & $\checkmark$ & $\checkmark$ &  & $\checkmark$  & & $\checkmark$ & & \\
				\hline 	
				\cite{biswas2021direct} & & $\checkmark$ & $\checkmark$ & & $\checkmark$  &\multicolumn{2}{c|}{Unspecified} &  & $\checkmark$ & $\checkmark$ \\
				\hline 	 	 					
				\cite{duan2019hybrid, guo2018exploiting, parks2014turbocharging, elmossallamy2019noncoherent, yang2017modulation} & & & $\checkmark$ &  & $\checkmark$ & $\checkmark$  & & & $\checkmark$ & $\checkmark$ \\
				\hline 	
				\cite{iyer2016inter, zhang2016enabling} & & & $\checkmark$ &\multicolumn{2}{c|}{Unspecified}  &\multicolumn{2}{c|}{Unspecified} & & $\checkmark$ & $\checkmark$ \\
				\hline 	 	
				\rowcolor{Gray2} \cite{kaplan2022direct}, this work & & $\checkmark$ & & $\checkmark$ & & $\checkmark$ & & & $\checkmark$ & $\checkmark$ \\
				\hline 			
		\end{tabular}}
	\end{table*}
	
	\subsubsection{Monostatic and Bistatic BC with Multiple-Antenna Technology}
	
	Here, we provide a literature summary on multiple-antenna technology for MoBC and BiBC, respectively.
	
	\textit{Monostatic:} In \cite{mishra2019optimal} and \cite{kashyap2016feasibility}, the authors show that the communication range of  monostatic BC systems increases when using multiple antennas at the reader. The authors of \cite{liu2014multi} maximize the minimum throughput for
	multiple battery-less single-antenna users in a MIMO monostatic system
	under a perfect channel state information (PCSI) assumption.
	
	In \cite{griffin2008gains} and \cite{boyer2013backscatter}, the authors prove that we can increase the diversity gain by increasing the number of antennas at the BD in a monostatic system under PCSI; consequently the  bit-error-rate (BER) performance is improved. 
	It is also shown in \cite{he2011closed} that the BER performance increases with an increasing number of BD antennas.
	
	\textit{Bistatic:} 
	In \cite{qu2022channel}, a channel estimation algorithm is proposed to estimate all the backscattering links in a multi-antenna CE and reader setup, and then a transceiver is designed using the estimates.
	In \cite{duan2017achievable}, a BD is added to a conventional multiple-input multiple-output (MIMO) system to convey additional information from the BD to a multi-antenna receiver applying joint decoding.
	
	\subsubsection{Interference Suppression in BC}
	In this subsection, we survey the literature on interference suppression in monostatic, ambient, and bistatic BC systems, respectively.
	
    \textit{Monostatic:}
	In \cite{villame2010carrier}, the authors propose a front-end architecture for the reader to cancel the self-interference (SI) in a full-duplex monostatic system. 
	In \cite{brauner2009novel}, the authors cancel the SI using a directional coupler and an adjustable reflective modulator. 
	The authors of \cite{bharadia2015backfi} propose a method to cancel the SI using the regenerated transmitted signal at the reader. 
	However,  interference cancellation methods for monostatic BC  are usually complex and have high  power consumption. They cannot directly be implemented in a BiBC system: the structure of the problem is different, and has new elements, such as a carrier frequency offset between the CE and reader \cite{tao2021novel}.
	
    \textit{Ambient:}
	Receive beamforming techniques to cancel the DLI are proposed for AmBC with a single-antenna transmitter setup in \cite{duan2019hybrid} and \cite{guo2018exploiting}. However, in \cite{duan2019hybrid}, the authors only provide the solution for a parametric channel model, and in \cite{guo2018exploiting}, DLI is canceled after the ADCs in the reader, which requires the use of high-resolution ADCs.	
	In \cite{parks2014turbocharging}, a multi-antenna receiver operating without digital computation is proposed to decode the backscattered signal by separating the DLI from the received signal. In \cite{iyer2016inter} and \cite{zhang2016enabling}, the authors avoid the DLI by shifting the carrier frequency of the incident signal in a BD in AmBC. However, the use of different frequency bands can reduce the spectral efficiency, and increase the power consumption and complexity at the BD. In \cite{elmossallamy2019noncoherent} and \cite{yang2017modulation}, the authors propose methods to avoid DLI in OFDM AmBC systems using  null subcarriers \cite{elmossallamy2019noncoherent} and cyclic prefixes \cite{yang2017modulation}. However, in \cite{yang2017modulation}, the interference is canceled after the ADC which does not solve the dynamic range problem. 
	
	\textit{Bistatic:}
	The authors of \cite{biswas2021direct} investigate the coverage region for IoT communication, and the effect of the DLI on the dynamic range in the BiBC and AmBC systems. They show that the high dynamic range limits the system performance.
	In \cite{varshney2017lorea} and \cite{li2019capacity}, the carrier frequency of the reflected signal is changed at the BD to solve the DLI problem in a single-input single-output (SISO) BiBC system. In \cite{tao2021novel}, the authors apply Miller coding at the BD and exploit the periodicity of the carrier signal to mitigate the DLI in a SISO BiBC system. However, the proposed method cancels the interference after the ADC which does not address the high-resolution ADC/high dynamic range problem. 	
	
	\subsection{Contributions and Organization}\label{sec:contributions}
	
	In this paper, we address the dynamic range problem and interference suppression for BiBC with a multi-antenna CE and reader. 
	To the best of our knowledge,  no previous work deals  with this setup.  
	Our specific contributions are:
	\begin{itemize}
		\item To address the high-resolution ADC/high dynamic range problem in a BiBC system, we propose a transmission scheme that suppresses the DLI by steering the transmission from the CE using beamforming.
		\item We analyze the effect of the proposed transmission scheme on the dynamic range under channel estimation errors.
		\item We derive an algorithm based on a generalized log-likelihood ratio test (GLRT) to detect the BD symbol/presence in BiBC. We propose an iterative algorithm to estimate the unknown parameters in the GLRT.
		\item Using the derived GLRT detector, we analyze the performance of BD symbol detection at the reader in a BiBC setup with multiple antennas.
	\end{itemize}
	
	Parts of the results in this paper were presented in \cite{kaplan2022direct}. The main difference between this paper and \cite{kaplan2022direct} is that herein all channels of a bistatic link are treated as unknown, whereas \cite{kaplan2022direct} assumes that the reader has perfect channel state information. As a result, novel algorithms for channel estimation are proposed in this paper which are necessary to detect the BD symbol/presence based on the GLRT. Note that the detection of the BD symbol and the presence of BD are the same problem. This is because we detect a pre-determined sequence of symbols from the BD in order to determine its presence as explained in detail in Section \ref{sec:detection_passive_device}.
	
	The remaining part of this paper is organized as follows. In Section \ref{sec:proposed_transmission_scheme}, we introduce the system model and our proposed transmission scheme. 
	In Section \ref{sec:channel_estimation_ineterference_supression}, we give the details of our interference suppression algorithm. 
	The GLRT detector and the estimation of unknown parameters are discussed in Section \ref{sec:detection_passive_device}. We present our simulation results in Section \ref{sec:numerical_results}. Finally, Section \ref{sec:conslision} concludes the paper.
	
	\textbf{Notation:} In this paper, $(\cdot)^T$, $(\cdot)^*$, and $(\cdot)^H$ denote transpose, conjugate, and Hermitian transpose, respectively. $\operatorname{Re}\{\cdot\}$ and $\operatorname{Tr}\{\cdot\}$ denote the real part of a signal and the trace of a matrix, respectively. 
	We use $\norm{\cdot}$ for Frobenius norm. 
	The cardinality of a set is denoted by $|\cdot|$.
	$E\{\cdot\}$ stands for the expected value. 
	$\operatorname{C}(\cdot)$ denotes the column space of a matrix.
	Italic, boldface capital, and boldface lowercase letters are used for scalars, matrices, and column vectors,  respectively.
	
	\section{Proposed Transmission Scheme} \label{sec:proposed_transmission_scheme}
	
	\begin{figure}[tbp]
		\centering
		\includegraphics[width = 0.9\linewidth]{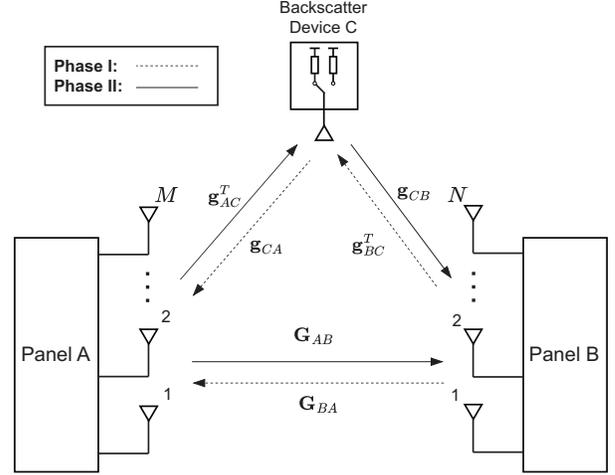}
		\caption{Model of the multiantenna BiBC system.}
		\label{fig:System_Model}
	\end{figure}
	
	In this section, we present a model of our bistatic communication system.
	We also describe our proposed transmission scheme, consisting of two phases: the channel estimation phase and the BD symbol detection phase.
	
	\subsection{System Model}  \label{sec:system_model}
	
	Fig. \ref{fig:System_Model} gives an overview of the system.
	Panel A (PanA) with $M$ antennas is the carrier emitter, Panel B (PanB) with $N$ antennas is the reader, and backscatter device C has a single antenna. The BD can change its antenna reflection coefficient by varying the impedance of the load connected to the antenna in order to modulate the backscattered signal. 	
	Additionally, PanA and PanB can be a part of a larger distributed MIMO setup with several panels \cite{van2019radioweaves}, and perform regular uplink/downlink communication, positioning, and sensing in addition to BC. Note, however, that if the  distributed MIMO system operates in time-division duplexing (TDD) mode (the default assumption, for example, in \cite{ngo2017cell}), then when communicating with the BD, one of the panels must switch to reception mode while the other panel is transmitting. In this respect, the interaction with the BD breaks the  TDD flow.
	
	Our aim is to decrease the required dynamic range of the reader, and detect the symbol/presence of the BD. 
	In Phase \RomanNumeralCaps{1} (P1), we estimate the channel between PanA and PanB. 
	In Phase \RomanNumeralCaps{2} (P2), we construct a beamformer using a projection matrix that is designed based on the estimated channel. The proposed beamformer decreases the dynamic range and increases the detection performance by suppressing the interference due to the direct link $\text{PanA} \rightarrow \text{PanB}$. \footnote{Note that our proposed interference suppression algorithm for BiBC uses beamforming in the CE, relying on explicit channel state information between the CE and the reader. This is feasible for the assumed BiBC setup, but it would not be possible for AmBC as there is no dedicated CE in AmBC.}
	
	In Fig. \ref{fig:System_Model}, $\mathbf{g}_{AC}^T, \mathbf{g}_{CA},  \mathbf{g}_{CB}, \mathbf{g}_{BC}^T, \mathbf{G}_{AB}$, and $\mathbf{G}_{BA}$ stand for the channels from PanA to BD, BD to PanA, BD to PanB, PanB to BD, PanA to PanB, and PanB to PanA, respectively. 
	Note that, channels are the effective baseband channels between the units, and not the wireless propagation channels. This is because, in its most general form, channels account for 1) wireless propagation effects, 2) (non-reciprocal) transceiver effects, and 3) calibration effects.
	In this paper, we have assumed that PanA and PanB are jointly reciprocity-calibrated such that
	$\textbf{G}_{AB}= \textbf{G}_{BA}^T$.	
	Here, the dimensions of $\textbf{g}_{AC}, \textbf{g}_{CA}, \textbf{g}_{CB}, \textbf{g}_{BC}$, and $\textbf{G}_{AB}$ are $M \times 1, M \times 1, N \times 1, N \times 1$, and $N \times M$, respectively. It is assumed that all channels are time-invariant during
	P1 and P2.
	
	\subsection{Transmission Scheme}
	
\begin{figure*}[tbp]				
		\centering		
		\includegraphics[width = 1\linewidth]{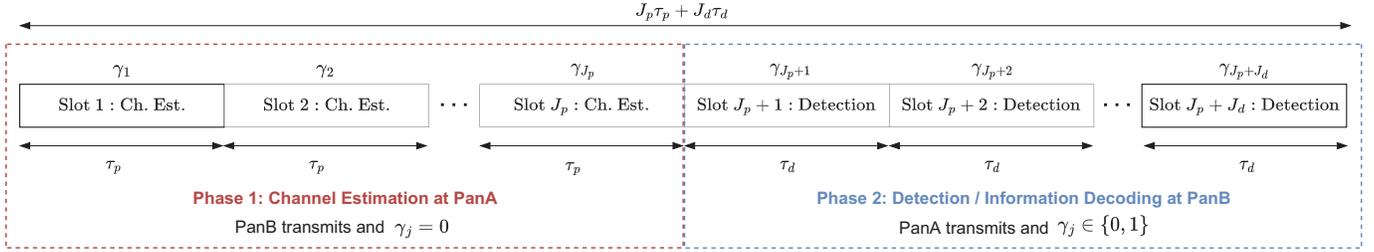}
		\caption{The proposed transmission scheme.}
		\label{fig:TransmissionScheme}
	\end{figure*}

	In Fig. \ref{fig:TransmissionScheme}, the proposed transmission scheme of our bistatic communication setup is illustrated. There are two phases, as explained below.

	\subsubsection{Phase \RomanNumeralCaps{1}: Channel Estimation at PanA}
	The first phase comprises $J_p$ slots ($\tau_p J_p$ symbols). 
	In each slot, PanB sends $N$ orthogonal pilot signals one per antenna, which each has $\tau_p$ symbols,  in order to facilitate estimation of $\mathbf{G}_{BA}$ at PanA. 
	The orthogonal pilot signals sent in a slot can be written in matrix form as $\boldsymbol{\Phi} \in \mathbb{C}^{N \times \tau_p}$ and satisfy
	\begin{equation}
		\boldsymbol{\Phi} \boldsymbol{\Phi}^{H} = \alpha_p \mathbf{I}_N,
	\end{equation}
	where $\alpha_p=\frac{p_t \tau_p}{N}$, $\tau_p \geq N$, and $p_t$ stands for the transmit power. The total transmitted energy during P1 is 
	\begin{equation}
		E_p  \triangleq J_p||\boldsymbol{\Phi}||^2 = J_p p_t \tau_p.
	\end{equation}
	
	In Fig. \ref{fig:TransmissionScheme}, $\gamma_j~\in~\{0,1\}$ denotes the reflection coefficient.
	In P1, we select $\gamma_j=0$ for $j\in \mathcal{S}_p =\{1,2,\dotsc,J_p\}$, i.e., the BD is silent in each slot. When the BD is silent, its reflection coefficient is a part of $\textbf{G}_{BA}$ like other scattering objects in the environment. (It is also possible to design a BD that absorbs the incoming signal for energy harvesting during $\gamma_j=0$ \cite{zawawi2018multiuser}; that, however, would not make a difference to our proposed algorithms.) When $\gamma_j=1$, the relative difference in the channel as compared to when $\gamma_j=0$ from PanB to PanA is $\textbf{g}_{CA} \textbf{g}_{BC}^T$.   Alternating between
	two different values of $\gamma_j$  corresponds to   on-off keying modulation with two states, which is commonly used also in much other literature, for example  \cite{hua2020bistatic, guo2018exploiting, zawawi2018multiuser}.
	
	\subsubsection{Phase \RomanNumeralCaps{2}: Detection at PanB}
	The second phase consists of $J_d$ slots ($\tau_d J_d$ symbols). In each slot, PanA sends a probing signal to detect the symbol of the BD at PanB. The probing signal sent in a slot can be represented in matrix form as $\boldsymbol{\Psi} \in \mathbb{C}^{M \times \tau_d}$ and satisfy 
	\begin{equation}
		\boldsymbol{\Psi} \boldsymbol{\Psi}^{H} = \alpha_d \textbf{I}_M,
	\end{equation} 
	where $\alpha_d = \frac{p_t \tau_d}{M}$ and $\tau_d \geq M$.

	\begin{table}[tbp]
	\caption{Summary of notation}
	\centering
	\label{tab:Parameters}
	\resizebox{0.46\textwidth}{!}{\begin{tabular}{|c|c|c|}
			\hline 
			\textbf{Parameter} & \textbf{Notation} & \textbf{Dimension} \\ \hline\hline
			Received signal at PanB & $\textbf{Y}_j$  & $N \times \tau_d$ \\
			\hline
			Received signal at PanA & $\textbf{Y}_j^p$  & $M \times \tau_p$ \\ 
			\hline
			Channel from PanA to PanB & $\textbf{G}_{A B}$  & $N \times M$ \\ \hline
			Channel from PanB to PanA & $\textbf{G}_{B A}$  & $M \times N$ \\ \hline
			Channel from PanA to BD & $\textbf{g}_{A C}^T$  & $1 \times M$ \\\hline
			Channel from BD to PanA & $\textbf{g}_{C A}$  & $M \times 1$ \\\hline
			Channel from BD to PanB & $\textbf{g}_{C B}$  & $N \times 1$ \\ \hline
			Channel from PanB to BD & $\textbf{g}_{B C}^T$  & $1 \times N$ \\ \hline
			Scaled projection matrix & $\textbf{P}_s$  & $M \times M$ \\ \hline
			Probing signal & $\boldsymbol{\Psi}$  & $M \times \tau_d$ \\ \hline
			Additive Gaussian noise at PanB & $\textbf{W}_j$  & $N \times \tau_d$ \\ 
			\hline
			Additive Gaussian noise at PanA & $\textbf{W}_j^p$  & $M \times \tau_p$ \\ \hline
			Pilot signal & $\boldsymbol{\Phi}$  & $N \times \tau_p$ \\
			\hline
			Reflection coefficient at BD & $\gamma_j$  & $1 \times 1$ \\ 
			\hline
			\multicolumn{3}{|c|}{Cardinalities: $|\mathcal{S}_p|=J_p, |\mathcal{S}_d|=J_d, |\mathcal{S}_d^0|=J_d^0$, and  $|\mathcal{S}_d^1|=J_d^1$}
			\\ 
			\hline
	\end{tabular}}
\end{table}
	
	The received signal of dimension $N \times \tau_d$ at PanB, in slot $j$ can be written as:
	\begin{equation} \label{eq:Phase2_InputOutputMatrixForm}
		\mathbf{Y}_j=\mathbf{G}_{A B}\textbf{P}_s \boldsymbol{\Psi}+\gamma_j \mathbf{g}_{C B} \mathbf{g}_{A C}^{T}  \textbf{P}_s \boldsymbol{\Psi} + \mathbf{W}_j,
	\end{equation}
	where $j\in \mathcal{S}_d=\{J_p+1,J_p+2,\dotsc,J_p+J_d\}$ and $J_p+J_d=J$. 	
	Similar to in P1, when the BD is silent ($\gamma_j=0$), its contribution to the propagation environment is considered to be included in $\textbf{G}_{AB}$. Therefore, $\mathbf{g}_{C B} \mathbf{g}_{A C}^{T}$ represents the \emph{difference} in the channel from PanA to PanB when $\gamma_j=1$ as compared to when $\gamma_j=0$. Due to this fact and because of reciprocity of propagation, $\textbf{G}_{AB}=\textbf{G}_{BA}^T$.	
	We assume that all   channels are time-invariant during the $J$ slot durations, i.e., the coherence time of all the channels exceeds $J_p \tau_p + J_d \tau_d$ symbols. 
	$\textbf{P}_s \in \mathbb{C}^{M \times M}$ is a scaled projection matrix introduced in order to minimize the DLI between PanA and PanB; the design principles for it will be explained in Section~\ref{sec:channel_estimation_ineterference_supression}. $\mathbf{W}_{j}~\in~\mathbb{C}^{N \times \tau_d}$ comprises additive Gaussian noise; all elements of $\textbf{W}_{j}$ are independent and identically distributed (i.i.d.) $\mathcal{CN}(0, 1)$.
	Depending on a pre-determined pattern associated  with the BD, in some slots of P1, $\gamma_j~=0$, and in the remaining slots $\gamma_j~=1$. 
	$\mathcal{S}_d$ is a set of $\{J_p+1,\dotsc,J\}$, and $\mathcal{S}_d^0$ and $\mathcal{S}_d^1$, which are subsets of $\mathcal{S}_d$, contain the indices for which $\gamma_j=0$ and $\gamma_j=1$ in P2, respectively. The cardinalities of  $\mathcal{S}_d, \mathcal{S}_d^0$, and  $\mathcal{S}_d^1$ are  
	$|\mathcal{S}_d|=J_d, |\mathcal{S}_d^0|=J_d^0$ and  $|\mathcal{S}_d^1|=J_d^1$, where $J_d=J_d^0+J_d^1$.
	The dimensions of the quantities in the model are summarized in Table \ref{tab:Parameters}.

	The next sections explain our proposed choice of the projector $\textbf{P}_s$ to minimize DLI,
	and the detection algorithm to be used at the reader. 
	
	\section{Proposed Interference Suppression Algorithm} \label{sec:channel_estimation_ineterference_supression}
	
	In this section, we first define the dynamic range in our system. Next, we present the channel estimation algorithm in P1 for the direct link between PanB and PanA. We then propose a novel algorithm based on the estimated direct link channel to mitigate the DLI at PanB in P2.
	The proposed algorithm decreases the required dynamic range of the system, and increases the SINR and the detection performance. In practice, it also enables the use of low-resolution ADCs due to the decreased dynamic range.
	
	\subsection{Dynamic Range}
	
	The dynamic range of an $n$-bit resolution ADC is $6.02n$ dB, and the quantization error of the ADC decreases exponentially with increasing $n$ \cite{baker2008cmos, lizon2020fundamentals}. 
		A high-resolution ADC, which has low quantization error, is required to detect the weak backscatter signal under strong DLI. 
		We also define the \emph{dynamic range of the received signal} during P2 as \cite{biswas2021direct, lietzen2020polarization}
	\begin{equation} \label{eq:DynamicRange} 
		\zeta = E\left\{\frac{\left\|\mathbf{G}_{A B} \textbf{P}_s \boldsymbol{\Psi}\right\|^{2}+\left\|\mathbf{g}_{C B} \mathbf{g}_{A C}^{T} \textbf{P}_s \boldsymbol{\Psi}\right\|^{2}}{\left\|\mathbf{g}_{C B} \mathbf{g}_{A C}^{T} \textbf{P}_s \boldsymbol{\Psi}\right\|^{2}}\right\},
	\end{equation}
	where the matrix product $\mathbf{G}_{A B}\textbf{P}_s \boldsymbol{\Psi}$ represents the DLI. In Eq. \eqref{eq:DynamicRange}, $\zeta$, the dynamic range of the received signal, is a good indicator of the required dynamic range of the reader circuitry.
	In this equation, the expectation is taken with respect to random  channel estimation errors (which affect $\textbf{P}_s$); the channels here are considered fixed. Note that the received signal from the BD is added to the numerator to satisfy $\zeta\ge 1$ (0 dB).
	When there is no projection, i.e., $\textbf{P}_s=\textbf{I}_M$, the dynamic range can be large:  $\zeta \gg 1$.
	When $\zeta \gg 1$, we need high-resolution ADCs which are not energy and cost efficient. This is because, in this operating regime, the backscattered signal is pushed to the last bits of the ADC, and the low-resolution ADCs cannot distinguish the weak backscatter signal under strong DLI due to the high quantization error.
	The aim of the scaled projection matrix, $\textbf{P}_s \in \mathbb{C}^{M \times M}$, is to project the transmitted signal onto the nullspace of the dominant directions of $\textbf{G}_{AB}$ (or more exactly, an estimate of it); consequently, the received DLI decreases which reduces the  dynamic range requirements on the reader circuitry.
	The design of the projection matrix is detailed in Section \ref{section:interferenceSuppression}.
	
	\subsection{Channel Estimation at PanA}
	
	In this subsection, we present the algorithm to estimate the channel from PanB to PanA, $\textbf{G}_{BA}$. 
	In the channel estimation phase, PanB sends the same pilot signal $\boldsymbol{\Phi}$ in each slot, that is, $J_p$ times. 
	At PanA, the received pilot signal in slot $j$ is given by
	\begin{equation} \label{eq:Phase1OriginalEquation}
		\textbf{Y}_j^p=\textbf{G}_{B A} \boldsymbol{\Phi} + \gamma_j \textbf{g}_{CA} \textbf{g}_{BC}^T \boldsymbol{\Phi} + \textbf{W}_j^p,
	\end{equation}
	where $j=1,2,\dots,J_p$ and $\textbf{W}_j^p \in \mathbb{C}^{M \times \tau_p}$ comprises additive noise and all elements of $\textbf{W}_j^p$ are i.i.d. $\mathcal{CN}(0,1)$. We select the reflection coefficients
	$\gamma_j=0$ for $j=1,2,\dotsc,J_p$, i.e., the BD is silent. As a result, Eq. (\ref{eq:Phase1OriginalEquation})  simplifies to
	\begin{equation}
		\textbf{Y}_j^p=\textbf{G}_{B A} \boldsymbol{\Phi} + \textbf{W}_j^p.
	\end{equation}
	
	The channel $\textbf{G}_{B A}$ is estimated by least-squares (LS) as follows:
	\begin{equation}
		\hat{\textbf{G}}_{B A}=\frac{1}{J_p} \sum_{j=1}^{J_p} \mathbf{Y}_j^p \boldsymbol{\Phi}^H (\boldsymbol{\Phi} \boldsymbol{\Phi}^H)^{-1}.
	\end{equation}
	Due to the reciprocity,  the channel $\textbf{G}_{AB}$ is simply $\textbf{G}_{AB}=\textbf{G}_{BA}^T$;
	the same holds for their estimates:  $\hat{\textbf{G}}_{A B}=\hat{\textbf{G}}_{B A}^T$. 
	
	The singular value decomposition (SVD) of $\hat{\textbf{G}}_{A B}$ can be written as
	\begin{equation}\label{eq:svd1}
		\hat{\textbf{G}}_{A B}=\textbf{U} \boldsymbol{\Delta} \textbf{V}^{\mathrm{H}},
	\end{equation}
	where $\textbf{U} \in \mathbb{C}^{N \times K_{0}}$ and $\textbf{V} \in \mathbb{C}^{M \times K_{0}}$ are semi-unitary matrices, and $K_{0} \leq \min\{M,N\}$ is the rank of $\hat{\textbf{G}}_{AB}$. $\boldsymbol{\Delta}$ is a $K_{0}\times K_{0}$ diagonal matrix with positive diagonal elements ordered in decreasing order. 
	
	\subsection{Interference Suppression Algorithm} \label{section:interferenceSuppression}
	
	This subsection explains our proposed design of the projector, $\textbf{P}_s$, whose application at PanA will reduce the DLI and consequently improve the detection performance at PanB.\footnote{Since the location  of the BD is unknown, we focus on the suppression of DLI, and we do not attempt to beamform power towards the BD in order to increase the backscattered power.} 
	
	In P2, PanA transmits a probing signal to enable PanB to  detect  the symbol of the BD. 
	The probing signal, $\boldsymbol{\Psi}$, satisfies $\boldsymbol{\Psi} \boldsymbol{\Psi}^{H}=\alpha_d \textbf{I}_M$. 
	Before transmitting $\boldsymbol{\Psi}$, PanA first designs a projection matrix based on the channel estimates obtained in P1. 
	After that, PanA projects the probing signal onto the nullspace of $\hat{\textbf{G}}_{AB}$ in order to minimize the  $\text{PanA} \rightarrow \text{PanB}$ DLI, decrease the dynamic range, and increase the detection probability of BD symbol at PanB.
	After the projection, PanA transmits the following signal
	\begin{equation} \label{eq:projection}
		\textbf{P}_s \boldsymbol{\Psi} = \Lambda \textbf{P} \boldsymbol{\Psi},
	\end{equation}
	where 
	$\textbf{P} $ is an orthogonal projection of dimension $M \times M$ and rank $M-K$, with $K$ to be appropriately selected, and
	\begin{equation}
		\Lambda=\sqrt{\frac{M}{M-K}}
	\end{equation}
	is the non-zero eigenvalue of  $\textbf{P}_s=\Lambda\textbf{P}$.
	In Eq. (\ref{eq:projection}), $\Lambda$ is used to keep the total radiated energy the same as without the projector, that is, 
	\begin{equation}
		E_d  \triangleq J_d||\boldsymbol{\Psi}||^2 = J_d||\textbf{P}_s\boldsymbol{\Psi}||^2.
	\end{equation} 
	
	We select $\textbf{P}$ to project onto the orthogonal complement of the  space spanned by the columns of $\textbf{V}_{K}$:
	\begin{equation}
		\textbf{P}=\textbf{I}-\textbf{V}_{K} \textbf{V}_{K}^{\mathrm{H}},
	\end{equation}
	where $\textbf{V}_{K}$ contains the first $K$ columns of $\textbf{V}$ in Eq. (\ref{eq:svd1}).
    The choice of the value of $K$ depends on several parameters, such as the number of antennas on the panels, the panel shapes (e.g., uniform linear arrays and rectangular linear arrays), the signal-to-noise ratio (SNR), and the channel model. For instance, when there are a strong line-of-sight (LoS) link and specular multipath components (SMC), setting $K$ to 1 makes it possible to cancel the LoS link and decrease the dynamic range. As $K$  increases, the dynamic range continues to decrease by the cancellation of SMCs, but the coverage area also decreases due to the increasing nulls in the antenna radiation pattern. Clearly, we must have $K \le M$. Selecting an appropriate value of $K$ is critical, and it could be chosen based on a predetermined dynamic range requirement and/or the number of dominant singular values of $\textbf{G}_{A B}$ (or more exactly, an estimate of it).

	Note that the location of the BD is unknown. To avoid reducing the backscatter link power, $\textbf{g}_{C B} \textbf{g}_{A C}^{T}$ should not lie in the 
	subspace spanned by the dominant right singular vectors of $\textbf{G}_{A B}$.
	For example, when $\operatorname{rank}(\textbf{G}_{AB})=1$ in line-of-sight conditions, the proposed algorithm requires that the CE, BD, and reader should not be located on a line, i.e., $ \textbf{g}_{C B} \textbf{g}_{A C}^{T}\neq \rho \textbf{G}_{A B}$
	for all $\rho \in \mathbb{C}$.
	
	In summary, the received signals during both phases are given by 
	\begin{equation} 
		\begin{aligned}
			&\textbf{Y}_j^p=\textbf{G}_{B A} \boldsymbol{\Phi} + \textbf{W}_j^p, j \in \{1,\dotsc,J_p\} \\
			&\textbf{Y}_j=\textbf{G}_{A B}\textbf{P}_s \boldsymbol{\Psi}+\gamma_j \textbf{g}_{C B} \textbf{g}_{A C}^{T}  \textbf{P}_s \boldsymbol{\Psi} + \textbf{W}_j, j \in \{J_p+1,\dotsc,J\}
			\label{eq:jam_coef_uncoded_1}
		\end{aligned}
	\end{equation} 
	
	\section{Detection of the Backscatter Device Symbol/Presence} \label{sec:detection_passive_device}
	
	In this section, we formalize the problem of detection during P2 as a hypothesis test, and develop a GLRT approach towards computing this test.
	The  hypothesis test models the two  scenarios absence and presence, respectively, of the BD  as $\mathcal{H}_{0}$ and $\mathcal{H}_{1}$.
	Specifically the test is:
	\begin{equation} \label{eq:hypothesisTesting}
		\begin{aligned}
			&\mathcal{H}_{0}: \begin{cases}\textbf{Y}_j=\textbf{G}_{A B} \textbf{P}_s \boldsymbol{\Psi} + \textbf{W}_j, & j \in \mathcal{S}_d \\
				\textbf{Y}_j^p=\textbf{G}_{B A} \boldsymbol{\Phi} + \textbf{W}_j^p, & j \in \mathcal{S}_p \end{cases} \\
			&\mathcal{H}_{1}: \begin{cases}\textbf{Y}_j =\textbf{G}_{A B} \textbf{P}_s \boldsymbol{\Psi} +\gamma_j \textbf{g}_{C B} \textbf{g}_{A C}^{T}  \textbf{P}_s \boldsymbol{\Psi}+ \textbf{W}_j, & j \in \mathcal{S}_d \\
				\textbf{Y}_j^p=\textbf{G}_{B A} \boldsymbol{\Phi} + \textbf{W}_j^p, & j \in \mathcal{S}_p. \end{cases}
		\end{aligned}
	\end{equation}
    Under $\mathcal{H}_{1}$, the BD varies its antenna reflection coefficient $\gamma_j$ according to a known (and pre-determined) pattern. Note that to detect the presence of the BD, we detect a pre-determined sequence of symbols from it. Therefore, technically, the problems of symbol detection and presence detection are equivalent. Furthermore, this hypothesis test can be interpreted as detecting the BD data, where the null hypothesis $\mathcal{H}_0$ corresponds to bit ``0'', and the alternative hypothesis $\mathcal{H}_1$ corresponds to bit ``1''.
	
	We consider the standard distributed MIMO setup, with a  backhaul link between the access points \cite[p. 32]{dmimo}, \cite{van2019radioweaves}. PanA and PanB can be any two antenna panels/access points in such a system.
	For this setup, we can assume that panels can share
	information, and also receive data over a backhaul network.
	Therefore, we assume that $\boldsymbol{\Psi}$, $\boldsymbol{\Phi}$, $\textbf{P}_s$, and $\textbf{Y}_j^p$ are known at PanB. The only information that needs to be sent through the backhaul network on a slot-by-slot timescale is $\textbf{P}_s$ and  $\textbf{Y}_j^p$. We also assume that $\textbf{G}_{A B}$, $\textbf{g}_{C B}$, and $\textbf{g}_{A C}$ are unknown by the receiver.
	
	The GLRT, using the received signals in P1 and P2, to detect the symbol/presence of the BD is given in Eq. \eqref{eq:GLRT_test}.	
	\begin{figure*}
		\begin{equation} \label{eq:GLRT_test}
			\frac{\max\limits_{\textbf{G}_{AB}, \mathbf{H}_{BL}} 
				\left(
				\prod\limits_{j\in\mathcal{S}_p} 
				p\left(\mathbf{Y}_{j}^p \mid \textbf{G}_{BA}\right)
				\prod\limits_{j\in\mathcal{S}_d} p\left(\mathbf{Y}_j \mid \mathcal{H}_{1}, \textbf{G}_{AB}, \mathbf{H}_{BL},  \gamma_j\right) 
				\right)}{\max\limits_{\textbf{G}_{AB}} 
				\left(
				\prod\limits_{j\in\mathcal{S}_p} 
				p\left(\mathbf{Y}_{j}^p \mid \textbf{G}_{BA}\right)
				\prod\limits_{j\in\mathcal{S}_d} 
				p\left(\mathbf{Y}_j \mid \mathcal{H}_{0}, \textbf{G}_{AB}\right)
				\right)}
			\underset{\mathcal{H}_{0}}{\overset{\mathcal{H}_{1}}{\gtrless}} \eta.
		\end{equation}
			\vspace{-12pt}	
	\end{figure*}
	In Eq. \eqref{eq:GLRT_test}, $p\left( \textbf{Y}_{j}^p \mid \textbf{G}_{BA} \right)$, $p\left(\mathbf{Y}_j \mid \mathcal{H}_{0}, \textbf{G}_{AB}\right)$, and $p\left(\mathbf{Y}_j \mid \mathcal{H}_{1}, \textbf{G}_{AB}, \mathbf{H}_{BL}, \gamma_j\right)$ denote the likelihood functions of the observations in P1, and under $\mathcal{H}_{0}$ and $\mathcal{H}_{1}$ in P2, respectively; they are given as follows:
	\begin{subequations} 
		\allowdisplaybreaks
		\begin{align}
			&p\left( \textbf{Y}_{j}^p \mid \textbf{G}_{BA} \right) = 
			\frac{1}{\pi^{M \tau_p}} \exp \left[-||\textbf{Y}_{j}^p-\textbf{G}_{B A} \boldsymbol{\Phi}||^2\right], \label{eq:pdf_P1}\\[6pt]
			&p\left(\mathbf{Y}_j \mid \mathcal{H}_{0}, \textbf{G}_{AB}\right) = \frac{1}{\pi^{N \tau_d}} \exp \left[-||\textbf{Y}_j-\textbf{G}_{A B} \textbf{P}_s \boldsymbol{\Psi}||^2\right], \\[6pt]
			&p\left(\mathbf{Y}_j \mid \mathcal{H}_{1}, \textbf{G}_{AB}, \mathbf{H}_{BL}, \gamma_j \right) \nonumber \\ 
			&\quad=\frac{1}{\pi^{N \tau_d}} \exp \left[-||\textbf{Y}_j-\textbf{G}_{A B} \textbf{P}_s \boldsymbol{\Psi} - \gamma_j \textbf{g}_{C B} \textbf{g}_{A C}^{T}  \textbf{P}_s \boldsymbol{\Psi}||^2\right]. \label{eq:norm_H1}
		\end{align}
			\vspace{-5pt}
	\end{subequations}
	The detection threshold is $\eta$, and 
	\begin{equation}
		\textbf{H}_{BL}=\textbf{g}_{C B} \textbf{g}_{A C}^{T} \textbf{P}_s
	\end{equation}  
	is an $N \times M$ rank-1 matrix, where $\textbf{g}_{C B} \textbf{g}_{A C}^{T}$ stands for the backscatter link cascade channel.

	The next subsections develop the estimate of $\textbf{G}_{AB}$ that maximizes the denominator of the the left hand side of Eq. (\ref{eq:GLRT_test}), assuming $\mathcal{H}_{0}$ is true, and the estimates of $\textbf{G}_{AB}$ and $\textbf{H}_{BL}$ that maximize the numerator in Eq. (\ref{eq:GLRT_test}), assuming $\mathcal{H}_{1}$ is true.
	
	\subsection{Estimation of Unknown Parameters Under $\mathcal{H}_{0}$}
	Under $\mathcal{H}_{0}$, $\textbf{G}_{AB}$ is the only unknown parameter. We estimate $\textbf{G}_{A B}$, that maximizes the denominator in Eq. (\ref{eq:GLRT_test}), using the received signals in P1 and P2 as follows: 
	\begin{subequations} \label{eq:H_DL_H_0_estimate}
		\allowdisplaybreaks
		\begin{align}
			\hat{\textbf{G}}_{A B} 
			&=\arg\max\limits_{\textbf{G}_{AB}}			 
			\prod_{j\in\mathcal{S}_p} 
			p\left(\mathbf{Y}_{j}^p \mid \textbf{G}_{BA} \right)
			\prod_{j\in\mathcal{S}_d} 
			p\left(\mathbf{Y}_j \mid \mathcal{H}_{0}, \textbf{G}_{AB}\right)
			\\
			&=\arg\min_{\textbf{G}_{A B}}  \label{eq:min_prob_1}
			\Biggl(  
			\sum_{j\in\mathcal{S}_p}
			||\mathbf{Y}_{j}^p - \textbf{G}_{B A} \boldsymbol{\Phi}||^2 \nonumber \\
			&\quad +
			\sum_{j\in\mathcal{S}_d}
			||\textbf{Y}_j - \textbf{G}_{A B} \textbf{P}_s \boldsymbol{\Psi}||^2
			\Biggl)\\
			&\stackrel{(a)}{=}\arg\min_{\textbf{G}_{A B}} \label{eq:min_prob_2}
			\Biggl(		
			\sum_{j\in\mathcal{S}_p}
			||\mathbf{Y}_{1,j} - \sqrt{\alpha_p}\textbf{G}_{AB} ||^2
		    \nonumber \\
		    &\quad +
			\sum_{j\in\mathcal{S}_d}
			||\textbf{Y}_{2,j} - \sqrt{\alpha_d}\textbf{G}_{A B} \textbf{P}_s||^2
			\Biggl)\\
			&\stackrel{(b)}{=}\arg\min_{\textbf{G}_{A B}} 
			\biggl( 	
			||\mathbf{Y}_{P1} - \sqrt{\alpha_p}\textbf{G}_{AB} \boldsymbol{\Upsilon}_{J_p}||^2
			\nonumber \\
			& \quad +
			||\textbf{Y}_{P2} - \sqrt{\alpha_d}\textbf{G}_{A B} \textbf{P}_s \boldsymbol{\Upsilon}_{J_d}||^2
			\biggl)\\    
			&=\arg\min_{\textbf{G}_{A B}} 
			\biggl(||\begin{bmatrix} \textbf{Y}_{P1} & \textbf{Y}_{P2} \end{bmatrix}  \nonumber \\
			&\quad - \textbf{G}_{A B} \begin{bmatrix} \sqrt{\alpha_p}\boldsymbol{\Upsilon}_{J_p}  & \sqrt{\alpha_d}\textbf{P}_s \boldsymbol{\Upsilon}_{J_d} \end{bmatrix} ||^2
			\biggl)\\    
			&=\begin{bmatrix} \textbf{Y}_{P1} & \textbf{Y}_{P2}  
			\end{bmatrix} 
			\begin{bmatrix} \sqrt{\alpha_p}\boldsymbol{\Upsilon}_{J_p}^T \\ \sqrt{\alpha_d}(\textbf{P}_s \boldsymbol{\Upsilon}_{J_d})^T \end{bmatrix} 
			\nonumber \\
			&\quad \left(
			\begin{bmatrix} \sqrt{\alpha_p}\boldsymbol{\Upsilon}_{J_p} & \sqrt{\alpha_d}\textbf{P}_s \boldsymbol{\Upsilon}_{J_d} \end{bmatrix} 
			\begin{bmatrix} \sqrt{\alpha_p}\boldsymbol{\Upsilon}_{J_p}^T \\ \sqrt{\alpha_d}(\textbf{P}_s \boldsymbol{\Upsilon}_{J_d})^T \end{bmatrix}
			\right)^{-1} \\ 
			&= \Biggl(   
			\sum_{j\in\mathcal{S}_p} \sqrt{\alpha_p} \textbf{Y}_{1,j} + 
			\sum_{j\in\mathcal{S}_d}      
			\sqrt{\alpha_d} \textbf{Y}_{2,j} \textbf{P}_s
			\Biggl) \nonumber \\
			&\quad
			(\alpha_p J_p \textbf{I}_M + \alpha_d J_d \textbf{P}_s)^{-1}
			\\
			&= \Biggl(    
			\sum_{j\in\mathcal{S}_p} \boldsymbol{\Phi}^* (\textbf{Y}_j^p)^T + 
			\sum_{j\in\mathcal{S}_d}     
			\textbf{Y}_j \boldsymbol{\Psi}^H \textbf{P}_s
			\Biggl) \nonumber \\
			&\quad
			(\alpha_p J_p \textbf{I}_M + \alpha_d J_d \textbf{P}_s)^{-1},
			\label{eq:H_DL_H_0_estimate_laststep}
		\end{align}
	\end{subequations}
	where equality $(a)$ is shown in Appendix \ref{FirstAppendix}.
	In \eqref{eq:min_prob_2}, $\textbf{Y}_{1,j}=\boldsymbol{\Phi}^*(\textbf{Y}_j^p)^T / \sqrt{\alpha_p}$ and $\textbf{Y}_{2,j}=\textbf{Y}_j \boldsymbol{\Psi}^H / \sqrt{\alpha_d}$. 
	In $(b)$, the block matrices $\textbf{Y}_{P1}\in \mathbb{C}^{N \times M J_p}$ and $\textbf{Y}_{P2}\in \mathbb{C}^{N \times M J_d}$ are created by ordering $\textbf{Y}_{1,j}$ and $\textbf{Y}_{2,j}$ in increasing order based on the index $j$, respectively. The block matrices $\boldsymbol{\Upsilon}_{J_p}\in \mathbb{R}^{M \times M J_p}$ and $\boldsymbol{\Upsilon}_{J_d}\in \mathbb{R}^{M \times M J_d}$ are $\boldsymbol{\Upsilon}_{J_p}=[\textbf{I}_M \cdots \textbf{I}_M]$ and $\boldsymbol{\Upsilon}_{J_d}=[\textbf{I}_M \cdots \textbf{I}_M]$. In Eq.~\eqref{eq:H_DL_H_0_estimate_laststep}, $(\alpha_p J_p \textbf{I}_M + \alpha_d J_d \textbf{P}_s)$ is invertible and positive definite.
	
	\vspace{-5pt}
	\subsection{Estimation of Unknown Parameters Under $\mathcal{H}_{1}$}
	
	Under $\mathcal{H}_{1}$, $\textbf{G}_{AB}$ and $\textbf{g}_{C B} \mathbf{g}_{A C}^{T}$ are the unknown parameters, but the estimate of $\textbf{g}_{C B} \mathbf{g}_{A C}^{T}$ is not unique when $\textbf{P}_s$ is not a full-rank matrix. Therefore, instead, we find the estimate of $\textbf{H}_{BL}$, which is unique. 
	
	We assume that reflection coefficients are known under $\mathcal{H}_{1}$. 
	We propose a cyclic optimization algorithm to  minimize the sum of the squared norms in Eqs. \eqref{eq:pdf_P1} and \eqref{eq:norm_H1}, and consequently to estimate  $\textbf{G}_{AB}$ and $\textbf{H}_{BL}$. The  algorithm consists of  following steps:
	\begin{itemize}
		\item \textit{Step 1:} First find an initial value by minimizing the objective function with respect to (w.r.t.) $\textbf{G}_{AB}$ using the observations $\mathbf{Y}_j$ and $\mathbf{Y}_{j^\prime}^p$, where $j \in \mathcal{S}_d^0$ and $j^\prime \in \mathcal{S}_{p}$. 
		\item \textit{Step 2:} Next,  minimize the objective function w.r.t. $\mathbf{H}_{BL}$ by using the estimated value of $\textbf{G}_{AB}$ and the observations $\mathbf{Y}_j$, where $j \in \mathcal{S}_{d}^1$. 
		As mentioned earlier, $\mathcal{S}_d^0$ and $\mathcal{S}_{d}^1$ denote the sets of reflection coefficient indices for which $\gamma_j=0$ and $\gamma_j=1$ in P2, respectively.
		\item \textit{Step 3:} Estimate $\textbf{G}_{AB}$ by using all observations and the estimated value of $\mathbf{H}_{BL}$.
		\item \textit{Step 4:} Iterate Steps 2--3 until convergence.
	\end{itemize}
	The details of these steps are given below.
	
	\subsubsection{Step 1 - Initial Estimation of $\textbf{G}_{AB}$} \label{sec:initial_estimation_G_AB}
	
	The estimate of $\textbf{G}_{AB}$ using the observations $\mathbf{Y}_j$ and $\mathbf{Y}_{j^\prime}^p$, where $j \in \mathcal{S}_d^0$ and $j^\prime \in \mathcal{S}_{p}$, is calculated (similar to Eq. (\ref{eq:H_DL_H_0_estimate}) but with $\mathcal{S}_d^0$ in lieu of $\mathcal{S}_d$) as
	\begin{subequations} 
		\vspace{-10pt}
		\label{eq:initial_estimate_G_AB_H1}
		\allowdisplaybreaks
		\begin{align}
			\hat{\textbf{G}}_{AB}^{\mathcal{H}_{1}} 
			&=\arg\min_{\textbf{G}_{AB}} 
			\biggl(  
			\sum_{j\in\mathcal{S}_p}
			||\mathbf{Y}_{j}^p - \textbf{G}_{B A} \boldsymbol{\Phi}||^2 \nonumber\\
			&\quad +
			\sum_{j\in\mathcal{S}_d^0} 
			||\textbf{Y}_j -  \textbf{G}_{A B} \textbf{P}_s \boldsymbol{\Psi} - \gamma_j \textbf{H}_{BL} \boldsymbol{\Psi}||^2 
			\biggl)\\
			&\stackrel{(a)}{=}\arg\min_{\textbf{G}_{AB}} \biggl( 
			\sum_{j\in\mathcal{S}_p}
			||\mathbf{Y}_{j}^p - \textbf{G}_{B A} \boldsymbol{\Phi}||^2 \nonumber\\
			&\quad +
			\sum_{j\in\mathcal{S}_d^0} 
			||\textbf{Y}_j -  \textbf{G}_{A B} \textbf{P}_s \boldsymbol{\Psi}||^2 
			\biggl) \\
			&= \biggl(   
			\sum_{j\in\mathcal{S}_p} \boldsymbol{\Phi}^* (\textbf{Y}_j^p)^T + 
			\sum_{j\in\mathcal{S}_d^0}      
			\textbf{Y}_j \boldsymbol{\Psi}^H \textbf{P}_s \biggl)
			\nonumber \\
			&\quad 
			(\alpha_p J_p \textbf{I}_M + \alpha_d J_d^0 \textbf{P}_s)^{-1}.
		\end{align}
	\end{subequations}
    In $(a)$, we have $\gamma_j=0$ for $j \in \mathcal{S}_d^0$.
	
	\subsubsection{ Step 2 - Estimation of $\mathbf{H}_{BL}$} \label{sec:estimation_of_h_bl}
	
	In this step, we minimize our objective function w.r.t. $\textbf{H}_{BL}$ by using $\hat{\textbf{G}}_{AB}^{\mathcal{H}_{1}}$ and the observations $\mathbf{Y}_j$, where $j \in \mathcal{S}_d^1$.
	
	To  estimate  $\textbf{H}_{BL}$, we apply the following steps. We first express the scaled projection matrix as 
	\begin{equation}
		\textbf{P}_s = \Lambda \textbf{P} = \Lambda \textbf{Q}\textbf{Q}^H
	\end{equation}
	in terms of its eigenvalue decomposition, where   $\Lambda=\sqrt{M/(M-K)}$ and $\textbf{Q}$ is an  $M \times (M-K)$ matrix that
	satisfies  $\textbf{Q}^H\textbf{Q}=\textbf{I}$.
	Note that $\operatorname{rank}{(\textbf{P}_s)}=M-K$.
	The minimization problem has two constraints: (1) $\textbf{H}_{BL}= \Lambda \textbf{g}_{C B} \mathbf{g}_{A C}^{T} \textbf{QQ}^H$ is a rank-1 matrix, and (2) $\textbf{H}_{BL}^H$ lies in the orthogonal complement of $\textbf{V}_{1,K}$, i.e., $\textbf{H}_{BL}^H \in \operatorname{C}(\textbf{P})$.
	To deal with this problem, we first estimate $\textbf{H}_{BL}^\prime = \textbf{g}_{C B} \mathbf{g}_{A C}^{T} \textbf{Q}$, which is an unconstrained rank-1 matrix, as follows:
	\begin{subequations} \label{eq:initial_g_cb}
		\allowdisplaybreaks
		\begin{align}
			\hat{\textbf{H}}_{BL}^\prime &=\arg\min\limits_{\textbf{H}_{BL}^\prime} \sum_{j \in \mathcal{S}_d^1} ||\textbf{Y}_j-\hat{\textbf{G}}_{AB}^{\mathcal{H}_{1}} \textbf{P}_s \boldsymbol{\Psi} - \gamma_j \Lambda \textbf{g}_{C B} \textbf{g}_{A C}^{T}  \textbf{P} \boldsymbol{\Psi}||^2 \\
			&\stackrel{(a)}{=}\arg\min\limits_{\textbf{H}_{BL}^\prime} ||\textbf{Y}_{DL} - \Lambda \textbf{g}_{C B} \textbf{g}_{A C}^{T}  \textbf{QQ}^H \textbf{D}||^2 \\
			&=\arg\min_{\textbf{H}_{BL}^\prime} 
			\big( ||\textbf{Y}_{DL}||^2+|| \Lambda \textbf{g}_{C B} \textbf{g}_{A C}^{T}  \textbf{QQ}^H \textbf{D}||^2 \nonumber\\
			&\quad 
			-2\operatorname{Re}\{\operatorname{Tr}\{\Lambda \textbf{D}^H \textbf{QQ}^H \textbf{g}_{A C}^{*} \textbf{g}_{C B}^H \textbf{Y}_{DL} \} \} \big)  \\
			&=\arg\min_{\textbf{H}_{BL}^\prime}
			\big( \operatorname{Tr}\{\Lambda^2 \textbf{g}_{CB} \textbf{g}_{AC}^T \textbf{QQ}^H \textbf{D} \textbf{D}^H \textbf{QQ}^H \textbf{g}_{A C}^{*} \textbf{g}_{C B}^H\} \nonumber\\
			&\qquad -2\operatorname{Re}\{\operatorname{Tr}\{ \Lambda \textbf{Y}_{DL} \textbf{D}^H \textbf{QQ}^H \textbf{g}_{A C}^{*} \textbf{g}_{C B}^H \} \} \big)  \\
			&=\arg\min_{\textbf{H}_{BL}^\prime}
			\big(J_d^1 \alpha_d \Lambda^2 ||\textbf{g}_{CB} \textbf{g}_{AC}^T \textbf{Q}||^2 \nonumber\\
			&\quad -2\operatorname{Re}\{\operatorname{Tr}\{ \Lambda \textbf{Y}_{DL} \textbf{D}^H\textbf{QQ}^H \textbf{g}_{A C}^{*} \textbf{g}_{C B}^H \} \} \big)  \\
			&=\arg\min_{\textbf{H}_{BL}^\prime}
			\bigg\{||\textbf{g}_{CB} \textbf{g}_{AC}^T \textbf{Q}||^2 \nonumber\\
			&\quad -
			\frac{2\operatorname{Re}\{\operatorname{Tr}\{ \textbf{Y}_{DL} \textbf{D}^H\textbf{QQ}^H \textbf{g}_{A C}^{*} \textbf{g}_{C B}^H \} \}}{J_d^1 \alpha_d \Lambda} \bigg\} \\
			&=\arg\min_{\textbf{H}_{BL}^\prime} \label{eq:h_bl_last_step}
			\left\| \textbf{g}_{CB} \textbf{g}_{AC}^T \textbf{Q}
			-\frac{1}{J_d^1 \alpha_d \Lambda} \textbf{Y}_{DL} \textbf{D}^H \textbf{Q} \right\|^2.
		\end{align}
	\end{subequations}
	In $(a)$, we have $\gamma_j=1$ for $j \in \mathcal{S}_d^1$, and the block matrix $\textbf{Y}_{DL}$ of dimension $N \times J_d^1 \tau_d$ is created by ordering $\{\textbf{Y}_j-\hat{\textbf{G}}_{AB}^{\mathcal{H}_{1}} \textbf{P}_s \boldsymbol{\Psi}\}$ in increasing order based on the index $j$.
	$\textbf{D}$   is an  $M \times J_d^1 \tau_d$-dimensional block matrix of $\boldsymbol{\Psi}$s: $\textbf{D}=[\boldsymbol{\Psi} \dotsc \boldsymbol{\Psi}]$, where $\textbf{D}\textbf{D}^H=J_d^1 \alpha_d \textbf{I}_M$. 
	
	We look for the best rank-one fit, in the  Frobenius-norm sense, in Eq. (\ref{eq:h_bl_last_step}). 
	The solution is given by the first term of the SVD of $\frac{1}{J_d^1 \alpha_d \Lambda}\textbf{Y}_{DL} \textbf{D}^H \textbf{Q}$ \cite{horn2012matrix}:
	\begin{equation} \label{eq:estimate_of_h_bl}
		\hat{\textbf{H}}_{BL}^\prime = \textbf{u}_{1} \delta_{1} \textbf{v}_{1}^H,
	\end{equation}
	where $\textbf{u}_{1}, \textbf{v}_{1}$, and $\delta_{1}$ are the dominant left singular vector, the dominant right singular vector, and the dominant singular value, respectively.
	
	Using Eq. (\ref{eq:estimate_of_h_bl}), the estimate of $\textbf{H}_{BL}$ that maximizes the numerator in the GLRT  is given by 
	\begin{equation} \label{eq:H_BL_under_H1}
		\hat{\textbf{H}}_{BL} =\Lambda \hat{\textbf{H}}_{BL}^\prime \textbf{Q}^H = \Lambda \textbf{u}_{1} \delta_{1} \textbf{v}_{1}^H \textbf{Q}^H,
	\end{equation}
	where $\hat{\textbf{H}}_{BL}^H$ lies in $\operatorname{C}(\textbf{P})$ because the eigenvectors of the projection matrix, i.e., the columns of $\textbf{Q}$, lie in $\operatorname{C}(\textbf{P})$.
	
	\subsubsection{Step 3 - Estimation of $\textbf{G}_{AB}$} 
	
	In Eq. (\ref{eq:initial_estimate_G_AB_H1}), the initial estimate of $\textbf{G}_{AB}$ is calculated without using the observations $\textbf{Y}_j$, where $j \in \mathcal{S}_d^1$. After the estimation of $\textbf{H}_{BL}$ in step 2, we can use all observations to estimate $\textbf{G}_{AB}$ (similar to Eq. (\ref{eq:H_DL_H_0_estimate}))  as follows:
	\begin{equation} \label{eq:estimated_G_AB_H1}
		\begin{aligned}
			\hat{\textbf{G}}_{AB}^{\mathcal{H}_{1}} &= \left(    
			\sum_{j\in\mathcal{S}_p} \boldsymbol{\Phi}^* (\textbf{Y}_j^p)^T + 
			\sum_{j\in\mathcal{S}_d}     
			(\textbf{Y}_j-\gamma_j \hat{\textbf{H}}_{BL} \boldsymbol{\Psi}) \boldsymbol{\Psi}^H \textbf{P}_s
			\right) \\
			&\quad (\alpha_p J_p \textbf{I}_M + \alpha_d J_d \textbf{P}_s)^{-1}.
		\end{aligned}
	\end{equation}
	
	\begin{algorithm}[tbp]
		\begin{algorithmic}[1]
			\REQUIRE $\textbf{Y}_j, \textbf{Y}_j^p, \gamma_j, \textbf{P}_s, \boldsymbol{\Psi}$
			\ENSURE $\hat{\textbf{H}}_{BL}, \hat{\textbf{G}}_{AB}^{\mathcal{H}_{1}}$
			\STATE $\upsilon=0$
			\STATE Compute \\ $\begin{aligned}
				\hat{\textbf{G}}_{AB}^{\mathcal{H}_{1}}(\upsilon) &= \left(             
				\sum\limits_{j\in\mathcal{S}_p} \boldsymbol{\Phi}^* (\textbf{Y}_j^p)^T + 
				\sum\limits_{j\in\mathcal{S}_d^0}     
				\textbf{Y}_j \boldsymbol{\Psi}^H \textbf{P}_s
				\right) \\
				&\quad (\alpha_p J_p \textbf{I}_M + \alpha_d J_d^0 \textbf{P}_s)^{-1} \end{aligned}$
			\STATE  Compute $\hat{\textbf{H}}_{BL}(\upsilon)=\Lambda \textbf{u}_{1}(\upsilon) \delta_{1}(\upsilon) \textbf{v}_{1}^H(\upsilon) \textbf{Q}^H$		
			\REPEAT 
			\STATE Compute \\ $\hat{\textbf{G}}_{AB}^{\mathcal{H}_{1}}(\upsilon+1)=
			\biggl(            
			\sum\limits_{j\in\mathcal{S}_p} \boldsymbol{\Phi}^* (\textbf{Y}_j^p)^T  + 
			\sum\limits_{j\in\mathcal{S}_d}     
			(\textbf{Y}_j$ \\ 
			$\quad -\gamma_j \hat{\textbf{H}}_{BL}(\upsilon) \boldsymbol{\Psi}) \boldsymbol{\Psi}^H \textbf{P}_s
			\biggl)
			(\alpha_p J_p \textbf{I}_M + \alpha_d J_d \textbf{P}_s)^{-1}.
			$
			\STATE  Compute $\hat{\textbf{H}}_{BL}(\upsilon+1)= \Lambda \textbf{u}_{1}(\upsilon+1) \delta_{1}(\upsilon+1) \textbf{v}_{1}^H(\upsilon+1) \textbf{Q}^H$
			\STATE $\upsilon = \upsilon +1$
			\UNTIL{$||\hat{\textbf{G}}_{AB}^{\mathcal{H}_{1}}(\upsilon)-\hat{\textbf{G}}_{AB}^{\mathcal{H}_{1}}(\upsilon-1)||^2   \leq \epsilon$}
		\end{algorithmic}
		\caption{The estimate of  $\textbf{G}_{AB}$ and $\textbf{H}_{BL}$ that maximize the numerator in the GLRT under $\mathcal{H}_{1}$.}
	\end{algorithm}	
	
	\subsubsection{Step 4 - Iteration} 
	
	Finally, we iteratively estimate $\textbf{H}_{BL}$ and $\textbf{G}_{AB}^{\mathcal{H}_{1}}$ by using Eqs. (\ref{eq:H_BL_under_H1}) and (\ref{eq:estimated_G_AB_H1}) until the Frobenius norm of the difference between two successive estimation of $\textbf{G}_{AB}^{\mathcal{H}_{1}}$ is smaller than a threshold. 
	
	A summary of our proposed algorithm is given in Algorithm~1. \footnote{Note that, if we assume that the pilot signal is sent from PanA to PanB, instead of from PanB to PanA in P1, the computational complexity of Algorithm~1 will remain the same.}
	
	\subsection{Modified Estimator for the Non-Jointly Calibrated Case}
	
 Throughout the above derivation, we have assumed that PanA and PanB are jointly reciprocity-calibrated such that
 $\textbf{G}_{AB}= \textbf{G}_{BA}^T$.
 In practice, such calibration can be achieved by distribution of a phase reference over a cable that connects the two panels, or by bi-directional intra- and inter- panel over-the-air measurements \cite{vieira2021reciprocity}.
 However, if such calibration has not been performed, and the panels are only \emph{individually} calibrated for reciprocity, there will be an unknown residual phase offset, say $\phi$, between the panels such that
 $\textbf{G}_{AB}=e^\phi \textbf{G}_{BA}^T$. In this scenario, the projection matrix designed in Section \ref{section:interferenceSuppression} will  cancel the DLI similar to when joint reciprocity calibration is assumed (since phase offset is only a complex scalar); however, we cannot directly use the signal received in P1 together with the signals received in P2 for the detection of the BD symbol.  A simple remedy in this case is to base the detector on \emph{only} the signals received during P2. The corresponding algorithm is worked out in  Appendix \ref{SecondAppendix}, and is principally different from Algorithm~1 developed above, as $\textbf{G}_{AB}$ is unidentifiable (even in the noise-free case) when only data from P2 are available; see the appendix for
 details.\footnote{In addition, it is also possible to use the modified estimator to decrease the backhaul overhead in the case of limited backhaul link capacity because there is no need to send the received signal in P1 over the backhaul link.}

	\subsection{Approximate GLRT Detector}
	Since $\hat{\textbf{G}}_{AB}^{\mathcal{H}_{1}}$ and $\hat{\textbf{H}}_{BL}$ are not maximum-likelihood estimates, inserting the estimates of $\textbf{G}_{AB}$ and $\textbf{H}_{BL}$ obtained above, we obtain an approximation of the GLRT detector in Eq.~(\ref{eq:GLRT_test}). 
	Specifically, the resulting (approximate) GLRT detector is given in Eq. \eqref{eq:GLRT_2}.	
	\begin{equation} \label{eq:GLRT_2}
	\resizebox{1\hsize}{!}{$ \begin{split}
			GLR &= \frac{    
				\prod\limits_{j\in\mathcal{S}_p} 
				p\left(\mathbf{Y}_{j}^p \mid \hat{\textbf{G}}_{AB}^{\mathcal{H}_{1}}\right)
				\prod\limits_{j\in\mathcal{S}_d} p\left(\mathbf{Y}_j \mid \mathcal{H}_{1}, \hat{\textbf{G}}_{AB}^{\mathcal{H}_{1}}, \hat{\textbf{H}}_{BL},  \gamma_j\right) 
			}{    
				\prod\limits_{j\in\mathcal{S}_p} 
				p\left(\mathbf{Y}_{j}^p \mid \hat{\textbf{G}}_{AB}^{\mathcal{H}_{0}}\right)
				\prod\limits_{j\in\mathcal{S}_d} 
				p\left(\mathbf{Y}_j \mid \mathcal{H}_{0}, \hat{\textbf{G}}_{AB}^{\mathcal{H}_{0}}\right)
			} \\
			& \underset{\mathcal{H}_{0}}{\overset{\mathcal{H}_{1}}{\gtrless}} \eta. \end{split}$}
	\end{equation}
	
	Defining $\textbf{A}_j, \textbf{B}, \textbf{C}_1$, and $\textbf{C}_2$ as follows,
	\begin{subequations}
		\allowdisplaybreaks
		\begin{align}
			\textbf{A}_j&=\hat{\textbf{G}}_{AB}^{\mathcal{H}_{1}} \textbf{P}_s \boldsymbol{\Psi} + \gamma_j \hat{\textbf{H}}_{BL} \boldsymbol{\Psi}, \\
			\textbf{B}&=\hat{\textbf{G}}_{AB}^{\mathcal{H}_{0}} \textbf{P}_s \boldsymbol{\Psi}, \\
			\textbf{C}_1 &= (\hat{\textbf{G}}_{AB}^{\mathcal{H}_{1}})^T \boldsymbol{\Phi}, \\
			\textbf{C}_2 &= (\hat{\textbf{G}}_{AB}^{\mathcal{H}_{0}})^T \boldsymbol{\Phi},
		\end{align}
	\end{subequations} 
	we can write $\log(GLR)$ as 
	\begin{subequations}
		\allowdisplaybreaks
		\begin{align}
			&\log(GLR) = -\sum_{j \in \mathcal{S}_d} ||\textbf{Y}_j-\hat{\textbf{G}}_{AB}^{\mathcal{H}_{1}} \textbf{P}_s \boldsymbol{\Psi} - \gamma_j \hat{\textbf{H}}_{BL} \boldsymbol{\Psi}||^2 
			\nonumber \\ 
			&\quad +\sum_{j \in \mathcal{S}_d} ||\textbf{Y}_j-\hat{\textbf{G}}_{AB}^{\mathcal{H}_{0}} \textbf{P}_s \boldsymbol{\Psi}||^2 
		    -\sum_{j \in \mathcal{S}_p} ||\textbf{Y}_j^p-(\hat{\textbf{G}}_{AB}^{\mathcal{H}_{1}})^T \boldsymbol{\Phi}||^2 
		    \nonumber \\
		    &\quad +\sum_{j \in \mathcal{S}_p} ||\textbf{Y}_j^p-(\hat{\textbf{G}}_{AB}^{\mathcal{H}_{0}})^T \boldsymbol{\Phi}||^2    
			\\
			&=-\sum_{j \in \mathcal{S}_d} \operatorname{Tr}\{(\textbf{Y}_j-\textbf{A}_j)(\textbf{Y}_j-\textbf{A}_j)^H\} 
			\nonumber \\
			&\quad
			 +\sum_{j \in \mathcal{S}_d} \operatorname{Tr}\{(\textbf{Y}_j-\textbf{B})(\textbf{Y}_j-\textbf{B})^H\} 
			\nonumber \\
			&\quad -\sum_{j \in \mathcal{S}_p} \operatorname{Tr}\{(\textbf{Y}_j^p-\textbf{C}_1)(\textbf{Y}_j^p-\textbf{C}_1)^H\} 
			\nonumber \\
			&\quad
			 +\sum_{j \in \mathcal{S}_p} \operatorname{Tr}\{(\textbf{Y}_j^p-\textbf{C}_2)(\textbf{Y}_j^p-\textbf{C}_2)^H\}
			\\
			&= \sum_{j \in \mathcal{S}_d} \big(2\operatorname{Re}\{\operatorname{Tr}\{ \textbf{Y}_j(\textbf{A}_j-\textbf{B})^H\}\}
			-||\textbf{A}_j||^2+||\textbf{B}||^2\big)
			\nonumber \\
			& \quad + \sum_{j \in \mathcal{S}_p} \big(2\operatorname{Re}\{\operatorname{Tr}\{ \textbf{Y}_j^p(\textbf{C}_1-\textbf{C}_2)^H\}\}
			-||\textbf{C}_1||^2+||\textbf{C}_2||^2\big).
		\end{align}
	\end{subequations}
	Finally, we can write the hypothesis test as follows:
	\begin{equation}
		\allowdisplaybreaks
		\begin{aligned}
			\label{eq:FinalGLRT}
			&\sum_{j \in \mathcal{S}_d} \left(2\operatorname{Re}\{\operatorname{Tr}\{ \textbf{Y}_j(\textbf{A}_j-\textbf{B})^H\}\}-||\textbf{A}_j||^2+||\textbf{B}||^2\right) \\
			&\quad +\sum_{j \in \mathcal{S}_p} \left(2\operatorname{Re}\{\operatorname{Tr}\{ \textbf{Y}_j^p(\textbf{C}_1-\textbf{C}_2)^H\}\}-||\textbf{C}_1||^2+||\textbf{C}_2||^2\right)
		 \\
			&\quad \underset{\mathcal{H}_{0}}{\overset{\mathcal{H}_{1}}{\gtrless}} \log(\eta).
		\end{aligned}
	\end{equation}
	
	\vspace{-7pt}
	\section{Numerical Results} \label{sec:numerical_results}
	
	In this section, we first provide the simulation parameters and then discuss numerical results. We assume that there is a specular reflector that causes a SMC. The channels are modeled as \footnote{In [1], in the numerical results, we used a far-field channel model which was not accurate given the wavelength and the distances in the setup. However, in this journal paper, we have addressed this issue by using the near-field channel model, as shown in Eq. (30). Note that, the far-field channel model and all the numerical results in [1] would remain the same and valid for a different sufficiently small value of $\lambda$ satisfying the far-field assumption.}
	\begin{subequations}
		\allowdisplaybreaks
		\begin{align}
			[\textbf{G}_{AB}]_{n,m} &= 	[\textbf{G}_{BA}^T]_{n,m} = \sqrt{\beta_{m,n}} e^{-j \frac{2\pi}{\lambda} d_{m,n}} \nonumber \\
			& \quad + g_{SMC}\sqrt{\beta_{m,n}^\prime} e^{-j \frac{2\pi}{\lambda} d_{m,n}^\prime}, \\
			[\textbf{g}_{AC}]_m &= [\textbf{g}_{CA}]_m = \sqrt{\beta_{m}} e^{-j \frac{2\pi}{\lambda} d_{m}} \nonumber \\
			&\quad + g_{SMC}\sqrt{\beta_{m}^\prime} e^{-j \frac{2\pi}{\lambda} d_{m}^\prime}, \\
			[\textbf{g}_{CB}]_n &= [\textbf{g}_{BC}]_n = \sqrt{\beta_{n}} e^{-j \frac{2\pi}{\lambda} d_{n}} \nonumber \\
			 &\quad + g_{SMC}\sqrt{\beta_{n}^\prime} e^{-j \frac{2\pi}{\lambda} d_{n}^\prime},
		\end{align}
	\end{subequations}
where $m\in \{1,2,\dotsc,M\}$, $n\in \{1,2,\dotsc,N\}$, $g_{SMC}$ is the amplitude gain of the SMC, $\lambda$ is the wavelength of the emitted signal, 
and $[\textbf{G}_{AB}]_{n,m}, 	[\textbf{g}_{AC}]_m$, and $[\textbf{g}_{CB}]_n$ are the $(n,m)^\text{th}$ element of $\textbf{G}_{AB}$,
 $m^\text{th}$ element of $\textbf{g}_{AC}$, and $n^\text{th}$ element of $\textbf{g}_{CB}$, respectively. The path-gain coefficients are defined as
	\begin{subequations}
		\allowdisplaybreaks
		\begin{align}
			\beta_{m,n}&= \frac{1}{d_{m,n}^2}, \beta_{m,n}^\prime= \frac{1}{(d_{m,n}^\prime)^2}, \\
			\beta_{m}&= \frac{1}{d_{m}^2}, \beta_{m}^\prime= \frac{1}{(d_{m}^\prime)^2}, \\
			\beta_{n}&= \frac{1}{d_{n}^2}, \beta_{n}^\prime= \frac{1}{(d_{n}^\prime)^2},
		\end{align}
	\end{subequations}
	where $d_{m,n}, d_{m},$ and $d_{n}$ stand for the LoS path lengths between the $m^\text{th}$ antenna in PanA - the $n^\text{th}$ antenna in PanB, the $m^\text{th}$ antenna in PanA - the BD, and the $n^\text{th}$ antenna in PanB - the BD, respectively. The non-LoS path lengths $d_{m,n}^\prime, d_{m}^\prime,$ and $d_{n}^\prime$ are defined similarly.
	We choose a uniform linear array at both PanA and PanB.
	
	Unless otherwise stated, we use the following parameters: $J_d=2, \tau_d=16, J_p=1, \tau_p~=~16,$ $M=16, N=16$, $\lambda=0.1 \text{ m}$, and $d_\text{ant}=0.5$, where $d_\text{ant}$ denotes the inter-antenna distance normalized by the carrier wavelength. We select the reflection coefficients at the BD as $\gamma_j = 0$ for $j \in \mathcal{S}_p$ in P1. In P2, $\gamma_j \in \{0,1\}$ for $j \in \mathcal{S}_d$, and we have the same number of $\gamma_j=0$ and $\gamma_j=1$, i.e., $|\mathcal{S}_d^0|=|\mathcal{S}_d^1|=J_d/2$. The SNR during the channel estimation phase is defined as $\text{SNR}_p=\bar{\beta}_{BA} p_t J_p \tau_p$, where $p_t$ is the transmit power and $\bar{\beta}_{BA}=\frac{\norm{\textbf{G}_{AB}}^2}{MN}$. The SNR during the detection of BD symbol is  $\text{SNR}_d=\bar{\beta}_{CB} \bar{\beta}_{AC} p_t J_d \tau_d \bar{\gamma}$, where $\bar{\gamma}=0.5$ is the average value of the reflection coefficients in P2, and $\bar{\beta}_{AC}=\norm{\textbf{g}_{AC}}^2/M$ and $\bar{\beta}_{CB}=\norm{\textbf{g}_{CB}}^2/N$. 
	The centers of PanA and PanB are located at positions with coordinates $(0,0)$ and $(6,0)$ in meters, respectively. The specular reflector is located along the x-axis at $y=-4$ m. Table \ref{tab:simulationParameters}   lists all  simulation parameters.
	
	\begin{table}[tbp]
		\caption{Simulation Parameters}
		\centering
		\label{tab:simulationParameters}
		\resizebox{0.485\textwidth}{!}{\begin{tabular}{|L|R|}	
				\hline 
				\textbf{Parameter} &  \textbf{Value} \\ \hline\hline			
				Number of slots for the probing signal & $J_d=2$ \\ \hline
				Number of symbols in each slot for the probing signal & $\tau_d=16$ \\ \hline
				Number of slots for the pilot signal & $J_p=1$ \\ \hline
				Number of symbols in each slot for the pilot signal  & $\tau_p=16$ \\ \hline			
				Number of antennas in PanA & $M=8,16$ \\ \hline
				Number of antennas in PanB & $N=8,16$ \\ \hline
				Number of antennas in BD & $1$ \\ \hline
				The location of PanA, PanB, and BD in meters & $(0,0), (6,0)$, and $(3,y),$ where $y \in [0\ 30]$  \\ \hline
				The location of the specular reflector in meters & $y=-4$  \\ \hline
				The amplitude gain of the SMC & $g_{SMC}=0.5$ \\ \hline
				Reflection coefficients at BD & $\gamma_j = 0$ for $j \in \mathcal{S}_p$ and $\gamma_j \in \{0,1\}$ for $j \in \mathcal{S}_d$ \\ \hline
				The wavelength of the emitted signal in meters & $\lambda=0.1$  \\ \hline
				Inter-antenna distance both in PanA and PanB in meter & $d_{\text{ant}}\lambda=0.5 \lambda=0.05$ \\ \hline
				SNR during the detection of BD symbol & $\text{SNR}_d=2 \text{ dB}$ \\
				\hline
				SNR during the channel estimation & $\text{SNR}_p=5, 20 \text{ dB}$ \\
				\hline			
		\end{tabular}}
	\end{table}
	
\begin{figure}[tbp]
		\centering
		\includegraphics[width = 0.8\linewidth] {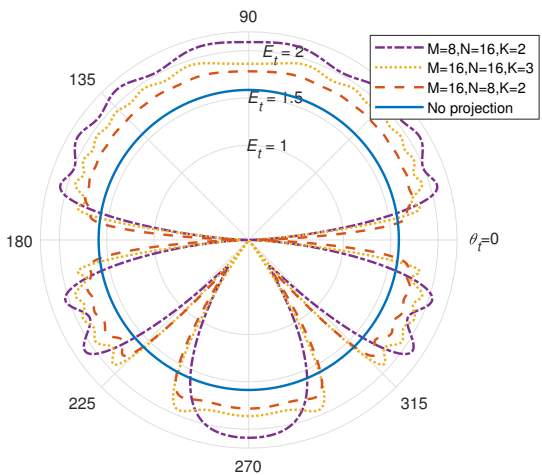}
		\caption{The 2-D antenna radiation patterns for the different number of antennas.}
		\label{fig:radiation_NM_8_16}
	\end{figure}

\begin{figure}[tbp]
		\centering
		\includegraphics[width = 0.8\linewidth] {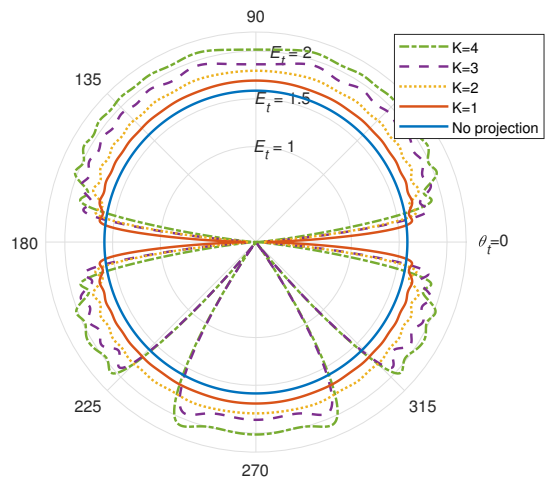}
		\caption{The 2-D antenna radiation patterns for the different $K$ values $(M=16,N=16)$.}
		\label{fig:radiation_NM_16}
	\end{figure}

	In Figs. \ref{fig:radiation_NM_8_16} and \ref{fig:radiation_NM_16}, we investigate the total radiated energy from PanA during a slot duration in P2, which is given by
	\begin{equation}
		E_t(\theta) \triangleq ||\textbf{g}(\theta)^T \textbf{P}_s \boldsymbol{\Psi}||^2=\frac{\alpha_d M}{M-K}||\textbf{g}(\theta)^T \textbf{P}||^2,
	\end{equation}
where $\textbf{g}(\theta) \in \mathbb{C}^{M \times 1}$ is the steering vector defined as follows:
\begin{equation} \label{eq:LoSchannel}
	\begin{aligned}
		\textbf{g}(\theta)= 
		\begin{bmatrix}
			1 \\
			\exp\left(j2\pi d_{\text{ant}}\sin(\theta) \right)  \\
			\vdots \\
			\exp\left(j(M-1) 2\pi d_{\text{ant}}\sin(\theta)\right)
		\end{bmatrix}
	\end{aligned},
\end{equation}
and $\theta$ is the angle of departure of the transmitted signal.
We select $\text{SNR}_d=2 \text{ dB}$, and assume that $\bar{\beta}_{AC}=1$ and $\bar{\beta}_{CB}=1$. The projection matrix is designed based on PCSI, i.e., $\hat{\textbf{G}}_{AB} = \textbf{G}_{AB}$, by PanA for all scenarios except the no-projection case, i.e., $\textbf{P}_s=\textbf{I}_M$. Note that the number of antennas in PanA and PanB does not affect the antenna radiation pattern for the no-projection~case.
	
	In Fig. \ref{fig:radiation_NM_8_16}, we compare the antenna radiation patterns for the three different cases: (1) $M=8, N=16$, (2) $M=16, N=16$, and (3) $M=16, N=8$. PanB is located at $0^{\circ}$, and in the first and third cases, the singular values of $\textbf{G}_{AB}$ are $(1.81, 0.59, 0.47, \dotsc)$, and we select $K=2$. In the second case, the singular values of $\textbf{G}_{AB}$ are $(2.33, 1.25, 0.79, 0.30, \dotsc)$, and we select $K=3$.
	As shown in the figure, the direct link, which is the LoS link plus SMC, is canceled due to the use of $\textbf{P}_s$ except for the no projection case. In addition, with increasing $M$, the beamforming accuracy and consequently the coverage area after projection increase in this particular setup.
	
	In Fig. \ref{fig:radiation_NM_16}, we compare the antenna radiation patterns for varying values of $K$ for $M=N=16$. 
	As seen in the figure, we can cancel more components of the direct link channel with increasing $K$. For example, the dominant directions of the LoS link are canceled for $K=1$ and $K=2$, and the dominant direction of the SMC is also canceled for $K=3$. This is because the first two dominant right singular vectors of $\textbf{G}_{AB}$ are mostly associated with the LoS link, while the third dominant right singular vector is mostly associated with the SMC. Note that the dominant $K$ right singular vectors of $\textbf{G}_{AB}$ are used for designing the projection matrix.
	
	\begin{figure}[tbp]
	\centering
	\includegraphics[width = 1\linewidth]{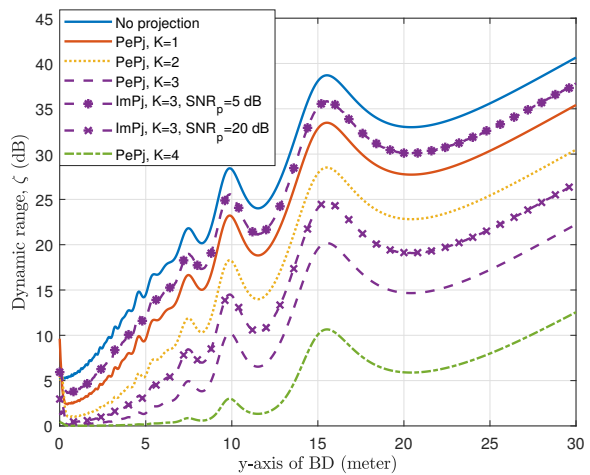}
	\caption{The dynamic range of the received signal.}
	\label{fig:DynamicRange_varyingK}
    \end{figure} 

	In Fig. \ref{fig:DynamicRange_varyingK}, we show the ratio $\zeta$ given in Eq. (\ref{eq:DynamicRange}) for different BD locations to investigate the change in the dynamic range in P2 for varying $K$ and $\text{SNR}_p$ values. The BD is located at $(3,y)$, where we change the vertical position of the BD, $y$, between $0$ and $30$ meters. For the perfect-projection (PePj) case, we assume that $\hat{\textbf{G}}_{A B} = \textbf{G}_{A B}$, for the no-projection case, $\textbf{P}_s=\textbf{I}_M$, and for the	imperfect-projection (ImPj) case, we design $\textbf{P}_s$ based on $\hat{\textbf{G}}_{A B}$.
	
    As seen in Fig. \ref{fig:DynamicRange_varyingK}, for $y=0$, the projection matrix cannot affect the dynamic range as intended since the BD is located on the broadside direction of the PanA. However, the dynamic range, $\zeta$, decreases with increasing $K$ when $y>0$ because we can cancel more components of the direct link channel. 
	For example, at $y=10 \text{ m}$, $\zeta$ is $28.32 \text{ dB}, 23.10\text{ dB}, 18.18\text{ dB}, 10.14\text{ dB},$ and $2.94\text{ dB}$ for the no projection, $K=1$, $K=2$, $K=3$ (PePj), and $K=4$ cases, respectively. 
	For $y<23$ m, there are fluctuations in the dynamic range curves due to the constructive and destructive interference between the LoS link and the SMC. However, after about 23 m, $d_m^\prime \approx d_m + c_1$ and $d_n^\prime \approx d_n + c_2$, and consequently $[\textbf{g}_{CB}]_n \approx c_3 e^{-j \frac{2\pi}{\lambda} d_{n}}$ and $[\textbf{g}_{AC}]_n \approx c_4 e^{-j \frac{2\pi}{\lambda} d_{m}}$, where $c_1,\dotsc,c_4$ are some constants. As a result, the channels $\textbf{g}_{CB}$ and $\textbf{g}_{AC}$ behave like a LoS channel without multipath. Therefore, the dynamic range curves are smoother beyond $y=23$ m for this particular setup.
	
    In Fig. \ref{fig:DynamicRange_varyingK}, we also show the effect of the imperfect projection on the dynamic range for $K=3$. As seen from the figure, the dynamic range, $\zeta$, decreases with increasing $\text{SNR}_p$ when $y>0$. 
	For example, at $y=10 \text{ m}$, $\zeta$ is $25.39 \text{ dB}$ at $\text{SNR}_p = 5$ dB and $14.38 \text{ dB}$ at $\text{SNR}_p = 20$ dB.
	This is because the projection matrix designed with high $\text{SNR}_p$ values has a better capability to decrease the DLI. 
	In practice, with a decreased dynamic range, it is possible to use low-resolution ADCs which are more cost and energy efficient than high-resolution ADCs.
	
	\begin{figure}[tbp]
		\centering
		\includegraphics[width = 1\linewidth]{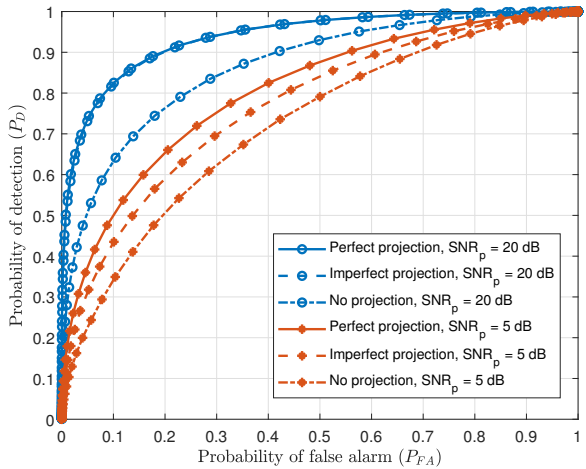}
		\caption{The BD symbol detection performance at PanB.}
		\label{fig:ROC}
	\end{figure}
	
	In Fig. \ref{fig:ROC}, simulation results for the hypothesis test in Eq. (\ref{eq:hypothesisTesting}) are shown. We use the GLRT detector in Eq. (\ref{eq:FinalGLRT}). 
	A triangular setup is used with the BD located at $(3,3)$ meters. We consider three different scenarios: (1) perfect projection, i.e., $\textbf{P}_s$ is designed based on $\hat{\textbf{G}}_{AB} = \textbf{G}_{AB}$, (2) imperfect projection, i.e., $\textbf{P}_s$ is designed based on the estimated channel, and (3) no projection, i.e.,  $\textbf{P}_s=\textbf{I}_M$. In all scenarios, we select $K=3$, $\text{SNR}_d=2 \text{ dB}$, and we use two different $\text{SNR}_p$ values as $5$ dB and $20$ dB in P1 to investigate the effect of the scaled projection matrix $\textbf{P}_s$ on the probability of detection $(P_D)$ and the probability of false alarm $(P_{FA})$ of the BD. Note that, although we do not estimate $\textbf{G}_{AB}$ using the pilot signal for the perfect projection and no projection cases, we send the pilot signal in P1 to use it in the detection phase (P2), for a fair comparison. 
	
	Compared to the system which works at low $\text{SNR}_p$ values, e.g., $5$ dB, the system which works at high $\text{SNR}_p$ values, e.g., $20$ dB, is superior in terms of projecting the transmitted signal onto the nullspace of the dominant direction of $\textbf{G}_{AB}$ because $\hat{\textbf{G}}_{A B}$ in P1 is more accurate at high $\text{SNR}_p$ values, and this affects the projection matrix accuracy. 
	As seen in the figure, the detection performance of the perfect and imperfect projection cases are almost the same at $\text{SNR}_p=20$ dB, while there is a performance difference between these two cases at $\text{SNR}_p=5$ dB.
	
	In addition, the performance of both the perfect and imperfect projection cases is better than that of the no projection case in each given $\text{SNR}_p$ value. This is because the radiated power in the directions which are close to the dominant directions of $\textbf{G}_{AB}$ decreases while the emitted power in all other directions increases due to the use of $\textbf{P}_s$. This phenomenon can also be seen in Fig.~\ref{fig:radiation_NM_16}. Compared to the no-projection case, for $K=3$ with the perfect projection case, the radiated power in all directions except the direct link is higher. For example, in Fig. \ref{fig:ROC} at $P_{FA}=0.1$, the imperfect projection case has almost $0.09$ and $0.19$ gain in the probability of detection when compared to the no projection case at $\text{SNR}_p=5$ dB and $\text{SNR}_p=20$ dB, respectively.
	
	\section{Conclusion} \label{sec:conslision}
	
	In this paper, we proposed a novel transmission scheme to be used in a bistatic BC setup with multiple reader and CE antennas. 
	The transmission scheme first estimates the channel between the CE (PanA) and the reader (PanB), and then uses this estimated channel to beamform the transmission from PanA. 
	In this beamforming, we propose to apply a specially designed  projection matrix whose effect is to  mitigate the DLI and decrease the required dynamic range of the reader circuitry.
	We showed that the dynamic range is significantly decreased by introduction of the  projection matrix in the beamforming, which in turn
	enables the use of  low-resolution ADCs which are low-cost and energy-efficient.
	Furthermore, we derive a detection algorithm based on a GLRT to detect the symbol/presence of the BD at the reader. 
	Joint usage of the proposed transmission scheme and detection algorithm results in a BiBC system with improved detection performance. This can be seen as a baseline approach to operate BiBC systems where both the CE and the reader are equipped with multiple antennas.

	\appendices
	
	\section{Proof of Proposition 1}
	\label{FirstAppendix}
	
	Since $\boldsymbol{\Psi} \boldsymbol{\Psi}^{H} = \alpha_d \textbf{I}_M$, the following two minimization problems are equivalent:
	\begin{equation} \label{eq:min_proof}
		\allowdisplaybreaks
		\begin{aligned} 
			&\arg\min_{\textbf{G}_{A B}}  
			||\textbf{Y}_j - \textbf{G}_{A B} \textbf{P}_s \boldsymbol{\Psi}||^2  \\
			&\quad =
			\arg\min_{\textbf{G}_{A B}}   ||\textbf{Y}_{2,j} - \sqrt{\alpha_d}\textbf{G}_{A B} \textbf{P}_s||^2.
		\end{aligned}
	\end{equation}
	To prove that, we first express
	\begin{subequations}
		\allowdisplaybreaks
		\begin{align}     
			& ||\textbf{Y}_j - \textbf{G}_{A B} \textbf{P}_s \boldsymbol{\Psi}||^2 = 
			||\mathbf{Y}_{j}||^2 + ||\textbf{G}_{A B} \textbf{P}_s \boldsymbol{\Psi}||^2 \nonumber \\
			&\quad -2\operatorname{Re}\{\operatorname{Tr}\{\mathbf{Y}_{j} \boldsymbol{\Psi}^H \textbf{P}_s \textbf{G}_{A B}^H \}\}, \\
			&||\textbf{Y}_{2,j} - \sqrt{\alpha_d}\textbf{G}_{A B} \textbf{P}_s||^2 = ||\textbf{Y}_{j} \boldsymbol{\Psi}^H / \sqrt{\alpha_d} - \sqrt{\alpha_d}\textbf{G}_{A B} \textbf{P}_s||^2 \nonumber \\
			&= \frac{||\textbf{Y}_{j} \boldsymbol{\Psi}^H||^2}{\alpha_d} + \alpha_d ||\textbf{G}_{A B} \textbf{P}_s||^2 -2\operatorname{Re}\{\operatorname{Tr}\{\textbf{Y}_{j} \boldsymbol{\Psi}^H \textbf{P}_s \textbf{G}_{A B}^H \}\}.
		\end{align}
	\end{subequations}
	Therefore,
	\begin{equation} 
		||\textbf{Y}_j - \textbf{G}_{A B} \textbf{P}_s \boldsymbol{\Psi}||^2 = 
		||\textbf{Y}_{2,j} - \sqrt{\alpha_d}\textbf{G}_{A B} \textbf{P}_s||^2 + c,
	\end{equation}
	where $c= ||\textbf{Y}_{j}||^2-||\textbf{Y}_{j} \boldsymbol{\Psi}^H||^2/\alpha_d$ depends on $\textbf{Y}_{j}$ and $\boldsymbol{\Psi}^H$, but not on $\textbf{G}_{AB}$, and hence does not effect the minimizer in Eq.~\eqref{eq:min_proof}. We can also apply the same calculation to $||\mathbf{Y}_{j}^p - \textbf{G}_{B A} \boldsymbol{\Phi}||^2$ and $||\mathbf{Y}_{1,j} - \sqrt{\alpha_p}\textbf{G}_{AB} ||^2$ since $\boldsymbol{\Phi} \boldsymbol{\Phi}^{H} = \alpha_p \mathbf{I}_N$.
	
	In addition, one can show that the elements of $\boldsymbol{\Phi}^*(\textbf{W}_j^p)^T / \sqrt{\alpha_p}$ and $\textbf{W}_j \boldsymbol{\Psi}^H / \sqrt{\alpha_d}$ matrices are i.i.d. $\mathcal{CN}(0,1)$. Therefore, the problems in Eqs. (\ref{eq:min_prob_1}) and (\ref{eq:min_prob_2}) are equivalent.

	\section{Estimation of Unknown Parameters}
	\label{SecondAppendix}
	
	When PanB does not know the received signal in P1, the hypothesis test becomes
	\begin{equation} \label{eq_ap:hypothesisTesting}
		\begin{split}
			\mathcal\mathcal\mathcal{H}_{0}&:  \textbf{Y}_j=\textbf{G}_{A B} \textbf{P}_s \boldsymbol{\Psi} + \textbf{W}_j\\
			\mathcal\mathcal\mathcal{H}_{1}&: \textbf{Y}_j =\textbf{G}_{A B} \textbf{P}_s \boldsymbol{\Psi} +\gamma_j \textbf{g}_{C B} \textbf{g}_{A C}^{T}  \textbf{P}_s \boldsymbol{\Psi}+ \textbf{W}_j,
		\end{split}
	\end{equation}
	where $j \in \mathcal{S}_d$. 
	
	The GLRT to detect the symbol/presence of the BD is as follows:
	\begin{equation} \label{eq_ap:GLRT_test2}
		\frac{\max\limits_{\mathbf{H}_{DL}, \mathbf{H}_{BL}} 
			\prod_{j\in\mathcal{S}_d} p\left(\mathbf{Y}_j \mid \mathcal{H}_{1}, \mathbf{H}_{DL}, \mathbf{H}_{BL},  \gamma_j\right)}{\max\limits_{\mathbf{H}_{DL}} 
			\prod_{j\in\mathcal{S}_d} p\left(\mathbf{Y}_j \mid \mathcal{H}_{0}, \mathbf{H}_{DL}\right)} \underset{\mathcal{H}_{0}}{\overset{\mathcal{H}_{1}}{\gtrless}} \eta,
	\end{equation}
	where
	\begin{equation}
		\textbf{H}_{DL}=\textbf{G}_{AB} \textbf{P}_s
	\end{equation}
	is an $N \times M$ matrix.
	
	Under $\mathcal{H}_{0}$, $\textbf{G}_{AB}$ is the unknown parameter, but the estimate of $\textbf{G}_{AB}$ is not unique because $\textbf{P}_s$ is not a full-rank matrix. We require, however, only an estimate of $\textbf{H}_{DL}=\textbf{G}_{A B} \textbf{P}_s$, rather than of $\textbf{G}_{A B}$, to find the maximum of the denominator in Eq.~(\ref{eq_ap:GLRT_test2}).
	
	We express the scaled projection matrix as 
	\begin{equation}
		\textbf{P}_s= \Lambda \textbf{P} = \Lambda \textbf{Q}\textbf{Q}^H.
	\end{equation}
	To  estimate  $\textbf{H}_{DL}$, we first estimate $\textbf{G}_{A B} \textbf{Q}$ as follows:
	\begin{subequations} \label{eq_ap:H_DL_H_0_estimate}
		\allowdisplaybreaks
		\begin{align}
			\widehat{\textbf{G}_{A B} \textbf{Q}}
			&=\arg\max\limits_{\textbf{G}_{AB} \textbf{Q}} 
			\prod_{j\in\mathcal{S}_d} p\left(\mathbf{Y}_j \mid \mathcal{H}_{0}, \textbf{H}_{DL}\right)\\
			&=\arg\min_{\textbf{G}_{A B}  \textbf{Q}} \sum_{j\in\mathcal{S}_d}
			||\textbf{Y}_j - \Lambda \textbf{G}_{A B} \textbf{Q}\textbf{Q}^H \boldsymbol{\Psi}||^2\\
			&=\arg\min_{\textbf{G}_{A B}  \textbf{Q}} \sum_{j\in\mathcal{S}_d}
			\big( ||\textbf{Y}_j||^2+||\Lambda \textbf{G}_{A B} \textbf{Q}\textbf{Q}^H \boldsymbol{\Psi}||^2 \nonumber \\
			&\quad
			-2\operatorname{Re}\{\operatorname{Tr}\{\Lambda \boldsymbol{\Psi}^H\textbf{QQ}^H\textbf{G}_{AB}^H \textbf{Y}_j \} \} \big)  \\
			&=\arg\min_{\textbf{G}_{A B}  \textbf{Q}}
			\sum_{j\in\mathcal{S}_d}
			\big( \operatorname{Tr}\{\Lambda^2 \textbf{G}_{AB} \textbf{QQ}^H \boldsymbol{\Psi} \boldsymbol{\Psi}^H \textbf{QQ}^H \textbf{G}_{AB}^H\} \nonumber \\
			&\quad -2\operatorname{Re}\{\operatorname{Tr}\{\Lambda \textbf{Y}_j \boldsymbol{\Psi}^H\textbf{QQ}^H\textbf{G}_{AB}^H \} \} \big)  \\
			&=\arg\min_{\textbf{G}_{A B}  \textbf{Q}}
			\sum_{j\in\mathcal{S}_d}
			\big(\Lambda^2 \alpha_d ||\textbf{G}_{A B}  \textbf{Q}||^2 \nonumber \\
			&\quad -2\operatorname{Re}\{\operatorname{Tr}\{\Lambda \textbf{Y}_j \boldsymbol{\Psi}^H\textbf{QQ}^H\textbf{G}_{AB}^H \} \} \big)  \\
			&=\arg\min_{\textbf{G}_{A B}  \textbf{Q}}
			\biggl\{J_d ||\textbf{G}_{A B}  \textbf{Q}||^2
			\nonumber \\
			&\quad
			 -2 \sum_{j\in\mathcal{S}_d}
			\frac{\operatorname{Re}\{\operatorname{Tr}\{\Lambda \textbf{Y}_j \boldsymbol{\Psi}^H\textbf{QQ}^H\textbf{G}_{AB}^H \} \}}{\Lambda^2 \alpha_d} \biggl\} \\
			&=\arg\min_{\textbf{G}_{A B}  \textbf{Q}}
			\left\| \textbf{G}_{A B}  \textbf{Q}
			-\frac{\sum_{j\in\mathcal{S}_d}\textbf{Y}_j \boldsymbol{\Psi}^H \textbf{Q}}{J_d \Lambda \alpha_d} \right\|^2 \\
			&=\frac{1}{J_d \Lambda \alpha_d}\sum_{j\in\mathcal{S}_d}\textbf{Y}_j \boldsymbol{\Psi}^H \textbf{Q}. \label{eq_ap:estimaeOfGabQ}
		\end{align}
	\end{subequations}
	We need to estimate $\textbf{H}_{DL} = \textbf{G}_{A B} \textbf{P}_s = \Lambda \textbf{G}_{A B} \textbf{QQ}^H$ subject to $\textbf{H}_{DL}^H \in \operatorname{C}(\textbf{P})$ under $\mathcal{H}_{0}$. Using Eq. (\ref{eq_ap:estimaeOfGabQ}), the estimate of $\textbf{H}_{DL}$ that maximizes the denominator in the GLRT  is given~by 
	\begin{equation} \label{eq_ap:H_DL_under_H0}
		\hat{\textbf{H}}_{DL}^{\mathcal{H}_{0}} = \widehat{\textbf{G}_{A B} \textbf{P}_s}=\Lambda \widehat{\textbf{G}_{A B} \textbf{Q}}\textbf{Q}^H = \frac{1}{J_d \alpha_d}\sum_{j\in\mathcal{S}_d}\textbf{Y}_j \boldsymbol{\Psi}^H \textbf{P},
	\end{equation}
	where $(\hat{\textbf{H}}_{DL}^{\mathcal{H}_{0}})^H$ lies in $\operatorname{C}(\textbf{P})$.
	
	The estimate of $\textbf{H}_{DL}$ under $\mathcal{H}_{1}$ needs to be calculated similar to Eqs. \eqref{eq_ap:H_DL_H_0_estimate} and \eqref{eq_ap:H_DL_under_H0} with small modifications. The estimation algorithm for $\textbf{H}_{BL}$ in Section \ref{sec:estimation_of_h_bl} can be directly used.
	
	\bibliographystyle{IEEEtran}
	\bibliography{references}
	
	\newpage

	\vspace{11pt}
	
	\begin{IEEEbiography}[{\includegraphics[width=1in,height=1.25in,clip,keepaspectratio]{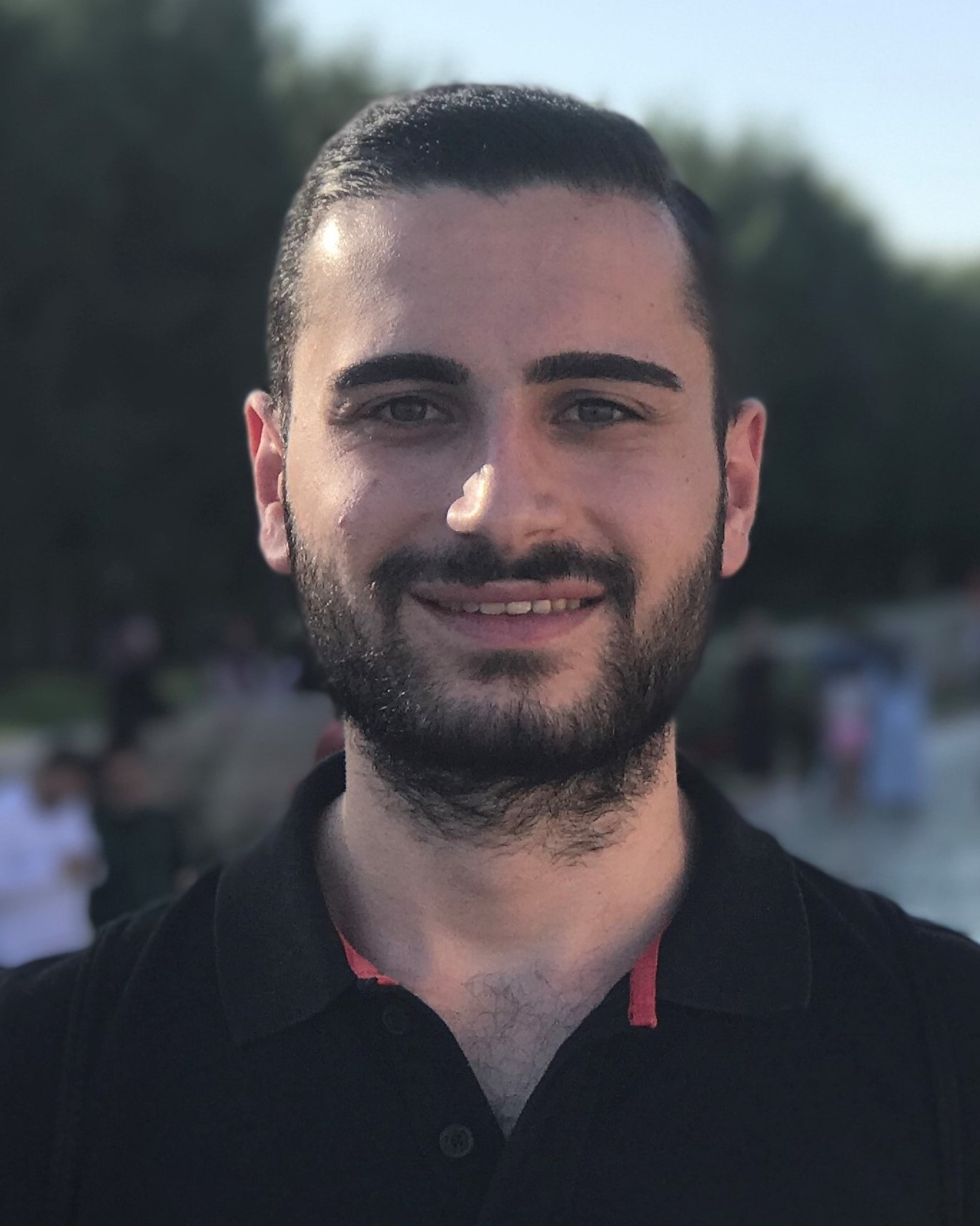}}]{Ahmet Kaplan}
		(S'20) was born in Istanbul, Turkey, in 1994. He received the B.Sc. and M.Sc. degrees (Hons.), in electronics and communication engineering, from the Istanbul Technical University, Istanbul, Turkey, in 2017 and 2020, respectively. From 2017 to 2019, he was a 5G Research Engineer with Turkcell, Istanbul, Turkey. He is currently a Research and Teaching Assistant with the Link\"oping University. His research interests include MIMO, backscatter communication, and low-density parity-check coding. He received the ICTC 2020 excellent paper award.
	\end{IEEEbiography}
	
	\vspace{11pt}
	
	\begin{IEEEbiography}[{\includegraphics[width=1in,height=1.25in,clip,keepaspectratio]{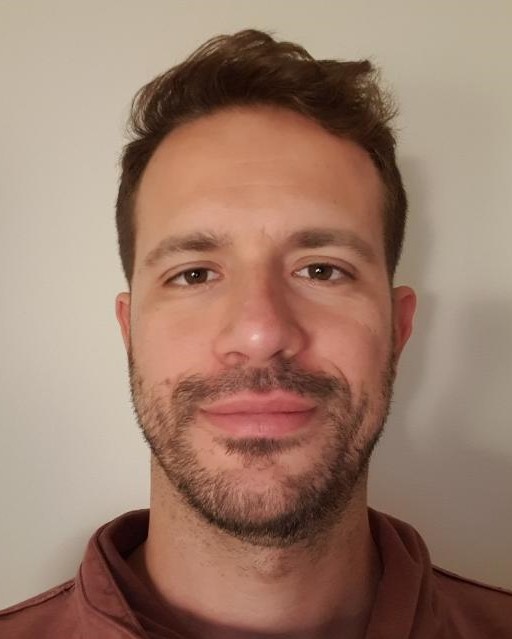}}]{Joao Vieira}
		 received the PhD degree from Lund University, Sweden, in 2017. Since then, he is with Ericsson Research investigating different 6G candidate technologies, especially those comprising large antenna arrays such as co-located and distributed massive MIMO.
	\end{IEEEbiography}
	
		\vspace{11pt}
	
	\begin{IEEEbiography}[{\includegraphics[width=1in,height=1.25in,clip,keepaspectratio]{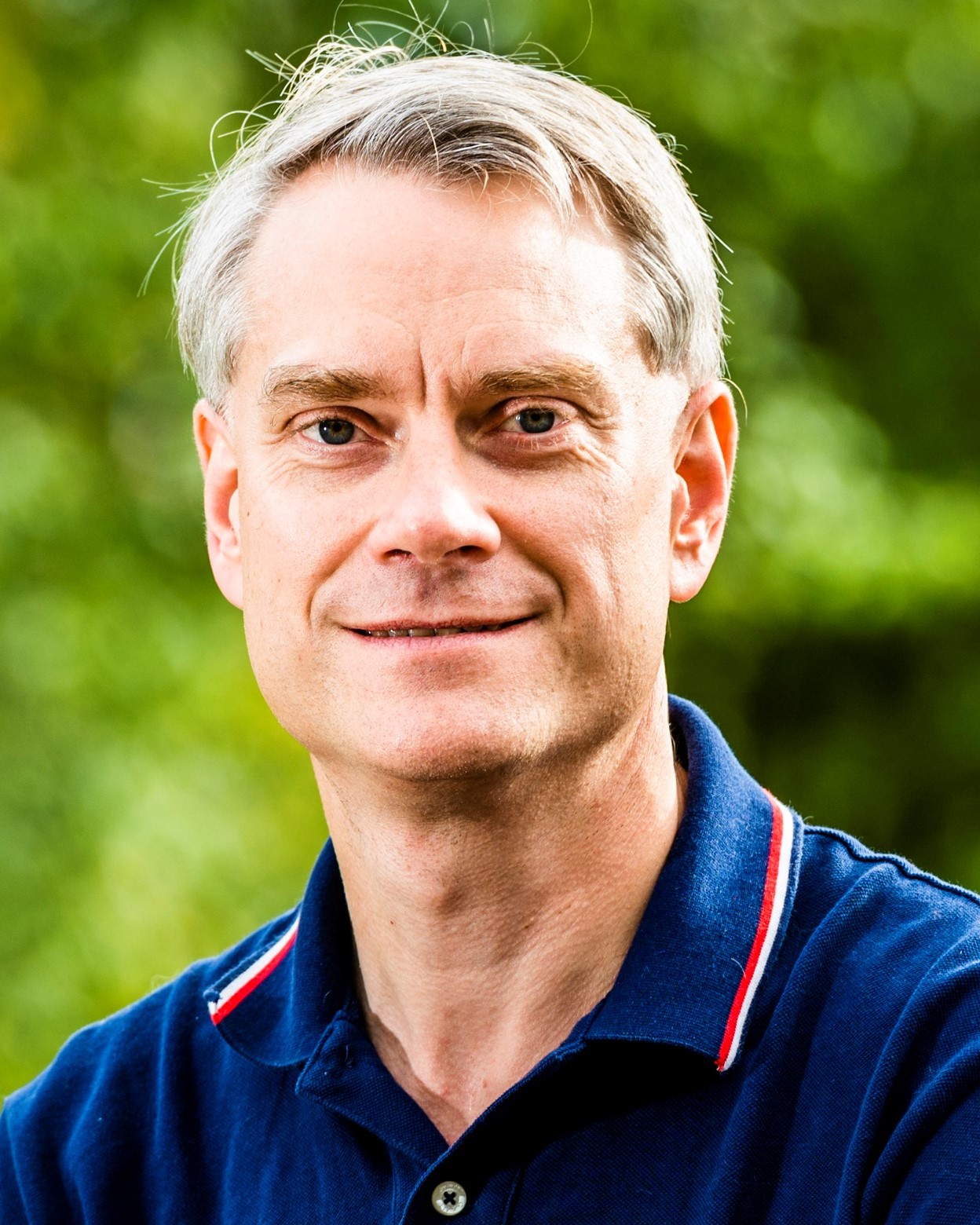}}]{Erik G. Larsson}
	(S'99--M'03--SM'10--F'16) received the Ph.D. degree from Uppsala University, Uppsala, Sweden, in 2002.  He is currently Professor of Communication Systems at Link\"oping University (LiU) in Link\"oping, Sweden. He was with the KTH Royal Institute of Technology in Stockholm, Sweden, the George Washington University, USA, the University of Florida, USA, and Ericsson Research, Sweden.  His main professional interests are within the areas of wireless communications and signal processing. He co-authored \emph{Space-Time Block Coding for  Wireless Communications} (Cambridge University Press, 2003) and \emph{Fundamentals of Massive MIMO} (Cambridge University Press, 2016).
	
	He served as  chair  of the IEEE Signal Processing Society SPCOM technical committee (2015--2016),
	chair of  the \emph{IEEE Wireless  Communications Letters} steering committee (2014--2015),
	member of the  \emph{IEEE Transactions on Wireless Communications}    steering committee (2019-2022),
	General and Technical Chair of the Asilomar SSC conference (2015, 2012),
	technical co-chair of the IEEE Communication Theory Workshop (2019),
	and   member of the  IEEE Signal Processing Society Awards Board (2017--2019).
	He was Associate Editor for, among others, the \emph{IEEE Transactions on Communications} (2010-2014),  the \emph{IEEE Transactions on Signal Processing} (2006-2010), and  the \emph{IEEE Signal  Processing Magazine} (2018-2022).
	
	He received the IEEE Signal Processing Magazine Best Column Award twice, in 2012 and 2014, the IEEE ComSoc Stephen O. Rice Prize in Communications Theory in 2015, the IEEE ComSoc Leonard G. Abraham Prize in 2017, the IEEE ComSoc Best Tutorial Paper Award in 2018, and the IEEE ComSoc Fred W. Ellersick Prize in 2019.
	\end{IEEEbiography}
	
	\vfill
	
\end{document}